\documentclass[a4paper,hidelinks, 12pt]{article}
\usepackage[utf8]{inputenc}
\usepackage{a4wide}
\usepackage{amsmath}
\usepackage{amsfonts}
\usepackage{amssymb}

\usepackage{subfigure}
\usepackage{cite}
\usepackage{xcolor}
\usepackage[toc,page]{appendix}
\numberwithin{equation}{section}
\usepackage{mathtools}
\usepackage{hyperref}
\usepackage{enumitem}
\def\lsim{\mathrel{\rlap{\raise 2.5pt \hbox{$<$}}\lower 2.5pt\hbox{$\sim$}}}

\allowdisplaybreaks

\newcommand{\hc}{{\rm h.c.}}

\def\gsim{\mathrel{\raise.3ex\hbox{$>$\kern-.75em\lower1ex\hbox{$\sim$}}}}

\begin{document}
\begin{titlepage}
\begin{center}

{\large \bf {Dark matter in a CP-violating three-Higgs-doublet model \\ with \boldmath$S_3$ symmetry}}

\vskip 1cm

A. Kun\v cinas, $^{a,}$\footnote{E-mail: Anton.Kuncinas@tecnico.ulisboa.pt}
O. M. Ogreid,$^{b,}$\footnote{E-mail: omo@hvl.no}
P. Osland$^{c,}$\footnote{E-mail: Per.Osland@uib.no} and 
M. N. Rebelo$^{a,}$\footnote{E-mail: rebelo@tecnico.ulisboa.pt}

\vspace{1.0cm}

$^{a}$Centro de F\'isica Te\'orica de Part\'iculas, CFTP, Departamento de F\'\i sica,\\ Instituto Superior T\'ecnico, Universidade de Lisboa,\\
Avenida Rovisco Pais nr. 1, 1049-001 Lisboa, Portugal,\\
$^{b}$Western Norway University of Applied Sciences,\\ Postboks 7030, N-5020 Bergen, 
Norway, \\
$^{c}$Department of Physics and Technology, University of Bergen, \\
Postboks 7803, N-5020  Bergen, Norway\\
\end{center}

\vskip 3cm

\begin{abstract}

In spite of the success of the Standard Model of Particle Physics, there are some theoretical predictions  which are not yet fully established experimentally as well as some experimental observations which cannot be fitted within its theoretical framework, thus requiring physics beyond the Standard Model. One of these is a hypothetical non-luminous form of matter -- dark matter. Models with an extended scalar electroweak sector yield plausible dark matter candidates. In this paper we study a specific model, C-III-a, from a family of $S_3$-symmetric three-Higgs-doublet models. The model consists of two active SU(2) doublets and an inert one. The latter is inert due to a $\mathbb{Z}_2$ symmetry that survives the breaking of $S_3$, and would accommodate a dark matter particle. We explore the model numerically, based on theoretical and experimental constraints. After applying a number of successive checks over the parameter space we found a viable dark matter mass region in the range $[6.5;\,44.5]~\text{GeV}$. This region is drastically different from the Higgs-like dark matter states that have been proposed: the well-known Inert Doublet Model and models with three scalar doublets, with one or two inert doublets. Furthermore, the C-III-a model allows for spontaneous CP violation. This means that the scalar potential explicitly conserves CP. However, in order to generate a realistic Cabibbo-Kobayashi-Maskawa matrix we need to introduce complex Yukawa couplings.

\end{abstract}

\end{titlepage}

%%%%%%%%%%%%%%%%%%%%%%%%%%%%%%%%%%%%%%%%%%%%%%%%
\section{Introduction}
%%%%%%%%%%%%%%%%%%%%%%%%%%%%%%%%%%%%%%%%%%%%%%%%

\begin{figure}[htb]
\begin{center}
\includegraphics[scale=0.50]{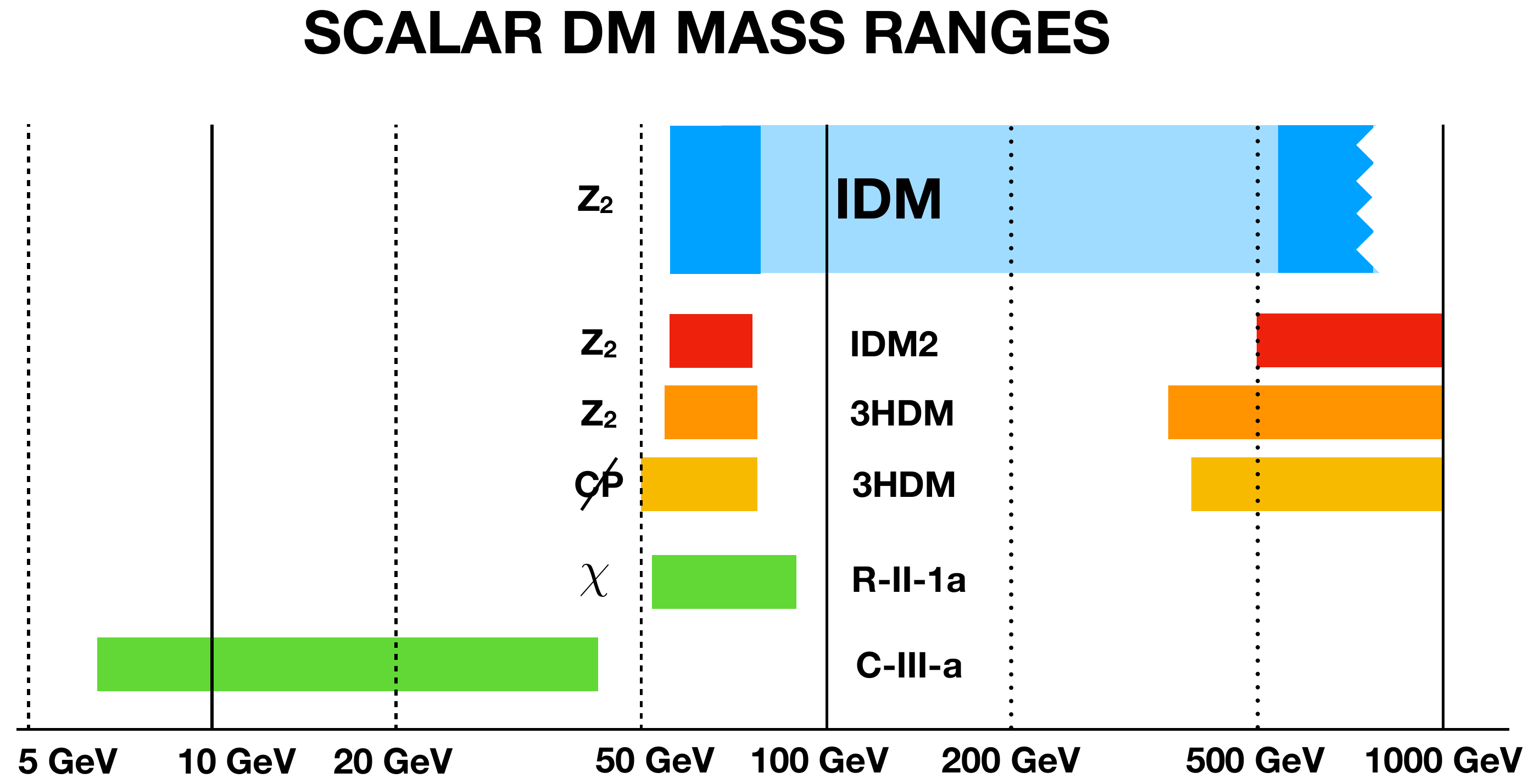}
\end{center}
\vspace*{-4mm}
\caption{ Sketch of allowed DM mass ranges up to 1~TeV in various models. Blue: IDM according to Refs.~\cite{Belyaev:2016lok,Kalinowski:2018ylg}, the pale region indicates a non-saturated relic density. Red: IDM2 \cite{Merchand:2019bod}. Ochre: three-Higgs-doublet model (3HDM) without \cite{Keus:2014jha,Keus:2015xya,Cordero:2017owj} and with CP violation \cite{Cordero-Cid:2016krd}. Green: $S_3$-symmetric 3HDM with a non-CP violating scalar sector (R-II-1a)~\cite{Khater:2021wcx} and with a CP violating scalar sector (C-III-a).}
\label{Fig:mass-ranges}
\end{figure}

A variety of models have been proposed in order to explain Dark Matter (DM), responsible for around a quarter of the total mass-energy density of the Universe~\cite{Planck:2018vyg}, in terms of scalar particles. The simplest models of this kind invoke an SU(2) singlet \cite{Silveira:1985rk,McDonald:1993ex} or an Inert Doublet Model (IDM) \cite{Deshpande:1977rw,Barbieri:2006dq}. Other models with additional SU(2) doublets have been proposed and studied. Among the latter, there are some in which the DM stability is provided by a remnant of the symmetry of the potential. Introducing additional SU(2) doublets, see figure~\ref{Fig:mass-ranges}, in general leads to more flexibility in accommodating dark matter:

\begin{enumerate}
\item
By having two non-inert doublets along with one inert doublet \cite{Grzadkowski:2009bt,Grzadkowski:2010au,Osland:2013sla,Merchand:2019bod,Khater:2021wcx}, which is the case studied here;
\item
By having one non-inert doublet along with two inert doublets \cite{Machado:2012ed,Keus:2013hya,Fortes:2014dca,Keus:2014jha,Aranda:2014lna,Keus:2015xya,Cordero-Cid:2016krd,Cordero:2017owj,Aranda:2019vda,Cordero-Cid:2020yba,Hernandez-Sanchez:2020aop}.
\end{enumerate}

%%%%%%%%%%%%%%%%%%%%%%%%%%%%%%%%%%%%%%%%%%%%%%%%
%%%%%%%%%%%%%%%%%%%%%%%%%%%%%%%%%%%%%%%%%%%%%%%%

Ideally, such models should also offer additional mechanisms for CP violation. An early model of this kind was the ``IDM2'' \cite{Grzadkowski:2009bt}. It builds on three SU(2) doublets, one of which is inert, whereas the two others basically constitute a CP-violating two-Higgs doublet model (2HDM)~\cite{Gunion:1989we,Branco:2011iw}. In the IDM2, the stability of the DM is provided by a $\mathbb{Z}_2$ symmetry that is imposed {\it ad hoc.}

In a companion paper \cite{Khater:2021wcx} we explored the possibility of having DM in models based on a spontaneously broken $S_3$ symmetry, and studied one of these models in detail. That model, denoted R-II-1a \cite{Emmanuel-Costa:2016vej}, does accommodate dark matter, but it has a real vacuum, and preserves CP. Here, we explore a rather similar model with real couplings, but with a complex vacuum, referred to as C-III-a, which violates CP spontaneously.

The paper is organised as follows. In section~\ref{sect:S3-potential} we introduce the $S_3$-symmetric potential, and discuss different dark matter candidates within the $S_3$-symmetric 3HDM. In section~\ref{sect:C-III-1a} the C-III-a model, on which the rest of our paper is based, is presented by giving the scalar masses, rotations leading to the physical scalars, scalar gauge couplings and the Yukawa couplings. It has been shown that the C-III-a model allows for spontaneous CP violation~\cite{Emmanuel-Costa:2016vej,Ogreid:2017alh}. In  section~\ref{sect:R-II-1a-vs-C-iii-c} we discuss similarities and differences between the \mbox{C-III-a} model and other models within the $S_3$-symmetric 3HDM. We discuss our approach to the numerical analysis of the model in section~\ref{sect:exp-constraints} by giving the model input and the theoretical and experimental constraints. The C-III-a model scan results are summarised in section~\ref{sect:Cut_3}. In section~\ref{sect:conclude} we present our conclusions.

%%%%%%%%%%%%%%%%%%%%%%%%%%%%%%%%%%%%%%%%%%%%%%%%%%%%
\section{The \boldmath$S_3$-symmetric models}
\label{sect:S3-potential}
\setcounter{equation}{0}
%%%%%%%%%%%%%%%%%%%%%%%%%%%%%%%%%%%%%%%%%%%%%%%%%%%%

%%%%%%%%%%%%%%%%%%%%%%%%%%%%%%%%%%%%%%%%%%%%%%%%%%%%
\subsection{The scalar potential}
%%%%%%%%%%%%%%%%%%%%%%%%%%%%%%%%%%%%%%%%%%%%%%%%%%%%

In terms of the $S_3$ singlet ($\mathbf{1}:h_S$) and doublet ($\mathbf{2}:\left(h_1\,~h_2\right)^\mathrm{T}$) fields, the $S_3$-symmetric potential can be written as \cite{Kubo:2004ps,Teshima:2012cg,Das:2014fea}:
\begin{subequations} \label{Eq:V-DasDey}
\begin{align}
V_2&=\mu_0^2 h_S^\dagger h_S +\mu_1^2(h_1^\dagger h_1 + h_2^\dagger h_2), \\
V_4&=
\lambda_1(h_1^\dagger h_1 + h_2^\dagger h_2)^2 
+\lambda_2(h_1^\dagger h_2 - h_2^\dagger h_1)^2
+\lambda_3[(h_1^\dagger h_1 - h_2^\dagger h_2)^2+(h_1^\dagger h_2 + h_2^\dagger h_1)^2]
\nonumber \\
&+ \lambda_4[(h_S^\dagger h_1)(h_1^\dagger h_2+h_2^\dagger h_1)
+(h_S^\dagger h_2)(h_1^\dagger h_1-h_2^\dagger h_2)+\hc] 
+\lambda_5(h_S^\dagger h_S)(h_1^\dagger h_1 + h_2^\dagger h_2) \nonumber \\
&+\lambda_6[(h_S^\dagger h_1)(h_1^\dagger h_S)+(h_S^\dagger h_2)(h_2^\dagger h_S)] 
+\lambda_7[(h_S^\dagger h_1)(h_S^\dagger h_1) + (h_S^\dagger h_2)(h_S^\dagger h_2) +\hc]
\nonumber \\
&+\lambda_8(h_S^\dagger h_S)^2.
\label{Eq:V-DasDey-quartic}
\end{align}
\end{subequations}
There are two coefficients in the potential that could be complex, thus CP can be broken explicitly. For simplicity, we have chosen all coefficients to be real. In spite of this choice there remains the possibility of breaking CP spontaneously. Notice that the $S_3$-symmetric potential, when written in terms of the irreducible representations, explicitly exhibits an inherent $\mathbb{Z}_2$ symmetry under which $h_1 \leftrightarrow - h_1$ (or equivalently $\{h_2,\,h_S\} \to - \{h_2,\,h_S\}$).

In the irreducible representation, the $S_3$ fields will be decomposed as
\begin{equation} \label{Eq:hi_hS}
h_i=\left(
\begin{array}{c}h_i^+\\ (w_i+\eta_i+i \chi_i)/\sqrt{2}
\end{array}\right), \quad i=1,2\,, \qquad
h_S=\left(
\begin{array}{c}h_S^+\\ (w_S+ \eta_S+i \chi_S)/\sqrt{2}
\end{array}\right),
\end{equation}
where the $w_i$ and $w_S$ parameters can be complex.

%%%%%%%%%%%%%%%%%%%%%%%%%%%%%%%%%%%%%%%%%%%%%%%%
\begin{figure}[htb]
\begin{center}
%\vspace*{-4mm}
\includegraphics[scale=0.4]{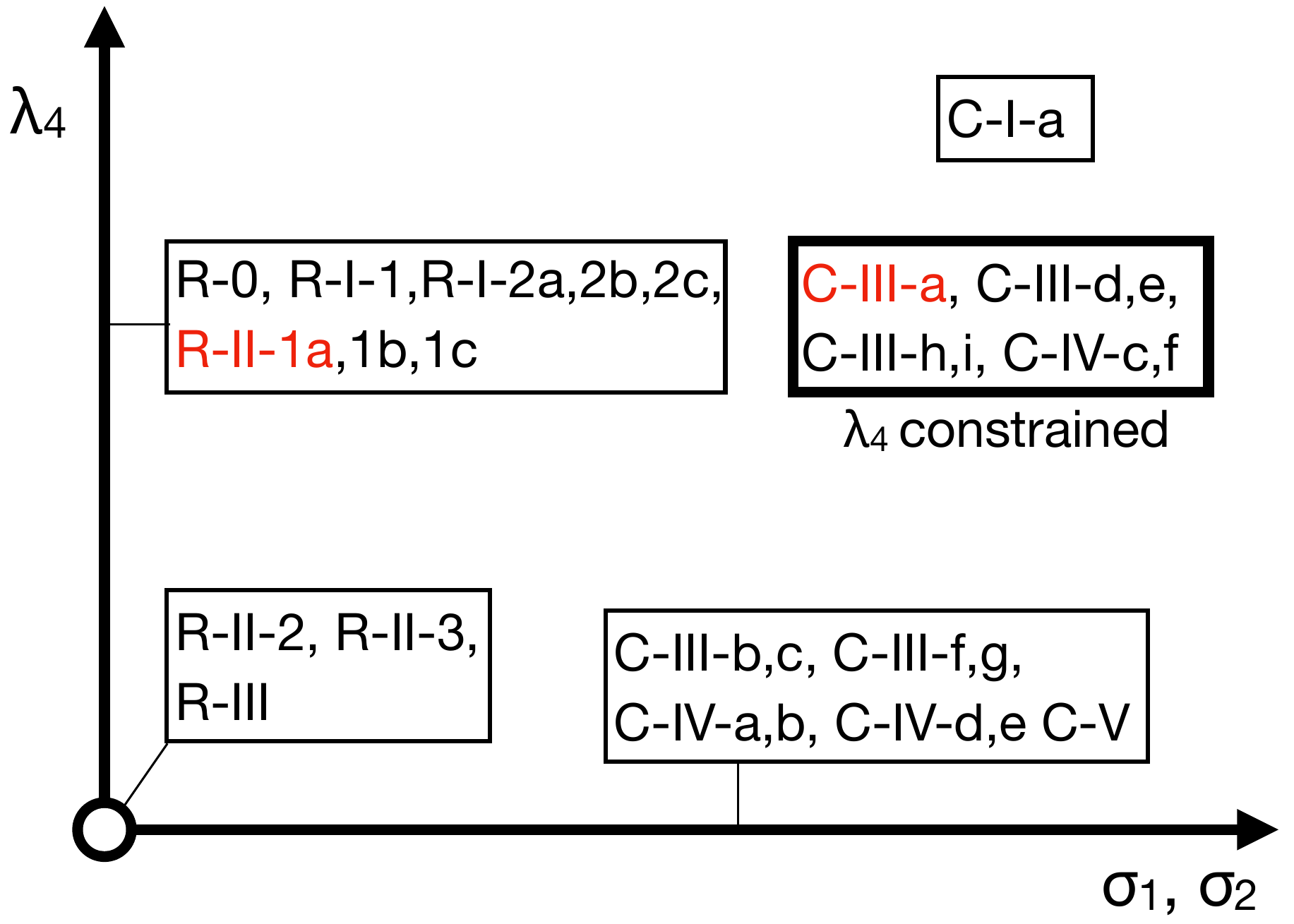}
\end{center}
\vspace*{-4mm}
\caption{ Overview of different vacua of the $S_3$-symmetric potential. The models in the heavy black box have $\lambda_4$ constrained by $\lambda_2+\lambda_3$ and/or $\lambda_7$. Continuous symmetries arise whenever $\lambda_4=0$. The model studied here and the one studied in the companion paper \cite{Khater:2021wcx} are indicated in red. The exact location of the boxes, other than the indication of whether or not they are on any of the axes or at the origin, is arbitrary.}
\label{Fig:lam4-sigma}
\end{figure}
%%%%%%%%%%%%%%%%%%%%%%%%%%%%%%%%%%%%%%%%%%%%%%%%

For the $S_3$-symmetric potential, 11 models with real vacuum expectation values (vevs), and 17 with at least one vev complex, have been identified~\cite{Emmanuel-Costa:2016vej}; different models correspond to different regions of parameter space. We list these models (vacua) in figure~\ref{Fig:lam4-sigma}, also indicating whether the vacuum is real (R-X-y) or complex (C-X-y). Our work will focus on the C-III-a model, which is an extension of the R-II-1a model~\cite{Khater:2021wcx}. Both of these models are highlighted in red in figure~\ref{Fig:lam4-sigma}. Along the horizontal axis $\sigma_1$ and $\sigma_2$ are the phases of $w_1$ and $w_2$ in the phase convention where $w_S$ is real.

The parameter $\lambda_4$ plays an important role. Soft symmetry-breaking terms are required whenever we work with solutions requiring $\lambda_4=0$, since in such cases most vacua lead to massless scalar states, Goldstone bosons, arising from the breaking of an O(2) symmetry. The symmetry of the potential can be softly broken by the following terms \cite{Kuncinas:2020wrn}:
\begin{equation}\label{VSoftGenericBasis}
\begin{aligned}
V_2^\prime =&\,\mu_2^2 \left( h_1^\dagger h_1 - h_2^\dagger h_2 \right) 
+ \frac{1}{2} \nu_{12}^2 \left( h_1^\dagger h_2 + \mathrm{h.c.} \right)\\&
+ \frac{1}{2} \nu_{01}^2 \left( h_S^\dagger h_1 + \mathrm{h.c.}\right)+ \frac{1}{2} \nu_{02}^2 \left( h_S^\dagger h_2 + \mathrm{h.c.} \right).
\end{aligned}
\end{equation}
In accordance with the previous simplification of couplings it is natural to assume that the soft terms are real. Although in this work we do not consider soft symmetry breaking some of the models presented (for completeness) in section~\ref{Sec:Different_DM_models} require soft terms.

%%%%%%%%%%%%%%%%%%%%%%%%%%%%%%%%%%%%%%%%%%%%%%%%
\subsection{The Yukawa interaction}
%%%%%%%%%%%%%%%%%%%%%%%%%%%%%%%%%%%%%%%%%%%%%%%%

Whenever the singlet vev, $w_S$, is different from zero we can construct a trivial Yukawa sector, ${\mathcal{L}_Y \sim 1_f \otimes 1_h}$ (subscripts ``$f$'' and ``$h$'' refer to fermions and scalars). In this case, the fermion mass matrices are:
\begin{subequations} \label{Eq.FMMws}
\begin{align}
{\cal M}_u=&\frac{1}{\sqrt{2}} \left( y_{ij}^u \right) w^*_S\,,\\
{\cal M}_d=&\frac{1}{\sqrt{2}} \left(y_{ij}^d \right )w_S\,,
\end{align}
\end{subequations}
where the $y$'s are the Yukawa couplings of the appropriate fermions and are not constrained by the $S_3$ symmetry. Therefore, in this case the Yukawa couplings are completely general.

Another possibility is when fermions transform non-trivially under $S_3$, with a Yukawa Lagrangian written schematically as ${\mathcal{L}_Y \sim (2 \oplus 1)_f \otimes (2 \oplus 1)_h}$, one doublet and one singlet of $S_3$,
\begin{equation*}
\mathbf{2}:\left(Q_1\,Q_2\right)^\mathrm{T},\,\left(u_{1R}\,u_{2R}\right)^\mathrm{T},\,\left(d_{1R}\,d_{2R}\right)^\mathrm{T}\quad\text{and}\quad\mathbf{1}:Q_3,\,u_{3R},\,d_{3R}.
\end{equation*}
Such structure yields the mass matrix for each quark sector ($d$ and $u$) of the form
\begin{subequations} \label{Eq:M}
\begin{align}
{\cal M}_u=
\frac{1}{\sqrt{2}}\begin{pmatrix}
y_1^uw_S^\ast+y_2^u w_2^\ast & y_2^u w_1^\ast & y_4^u w_1^\ast \\
y_2^u w_1^\ast & y_1^u w_S^\ast-y_2^u w_2^\ast & y_4^u w_2^\ast \\
y_5^u w_1^\ast & y_5^u w_2^\ast & y_3^u w_S^\ast
\end{pmatrix},\\
{\cal M}_d=
\frac{1}{\sqrt{2}}\begin{pmatrix}
y_1^dw_S+y_2^dw_2 & y_2^d w_1 & y_4^d w_1 \\
y_2^d w_1 & y_1^d w_S-y_2^d w_2 & y_4^d w_2 \\
y_5^d w_1 & y_5^d w_2 & y_3^d w_S
\end{pmatrix}.
\end{align}
\end{subequations}

Let us briefly consider what happens with the Yukawa sector in this case. When the DM candidate resides in the scalar $S_3$ singlet, $w_S = 0$, we need the fermions only to couple to the $S_3$ doublet, schematically represented by ${\mathcal{L}_Y \sim (2 \oplus 1)_f \otimes 2_h}$. Another possibility is when the DM candidate resides in the scalar $S_3$ doublet. To keep notation simple, we shall write the Yukawa sector ${\mathcal{L}_Y \sim (2 \oplus 1)_f \otimes (2 \oplus 1)_h}$, assuming that $w_S \neq 0$, as the general form of the fermion mass matrices persists. However, in order to stabilise the DM candidate one needs to introduce an additional $\mathbb{Z}_2$ symmetry in the Yukawa sector to decouple a specific inert doublet from the fermionic sector. Notice that whenever $w_1=0$ which is the case in the model we study, the mass matrices become block-diagonal. This case does not generate a realistic CKM matrix. Therefore, we shall require that the quarks transform trivially under $S_3$ which means that they can only couple to the $S_3$-singlet Higgs doublet.

We recall that for a scalar doublet to accommodate a DM candidate it must have a vanishing vev, since otherwise it would decay via its gauge couplings (e.g., the $SW^+W^-$ and $SZZ$ couplings).  Such requirement puts severe restrictions on the Yukawa interactions: as the number of free parameters, dependent on the vev, is reduced, it gets complicated to generate realistic fermionic masses and a complex Cabibbo-Kobayashi-Maskawa (CKM) matrix. In some cases realistic quark masses and mixing can only be generated if the quarks are taken to be $S_3$ singlets and only couple to the $h_S$ doublet.

%%%%%%%%%%%%%%%%%%%%%%%%%%%%%%%%%%%%%%%%%%%%%%%%
\subsection{Dark matter candidates in \boldmath$S_3$-based 3HDM}\label{Sec:Different_DM_models}
%%%%%%%%%%%%%%%%%%%%%%%%%%%%%%%%%%%%%%%%%%%%%%%%

Some of the $S_3$-symmetric models~\cite{Emmanuel-Costa:2016vej} have vacua minimised for $\lambda_4=0$. Such models are associated with unwanted, additional, Goldstone bosons. Soft breaking terms of the $S_3$ symmetry would have to be introduced in the potential~\cite{Kuncinas:2020wrn}, note that soft breaking is not possible in the Yukawa sector. When introducing soft breaking terms, constraints will change. However, we will retain the nomenclature of the unbroken case from which they originate, thus when adding soft-breaking terms to R-I-1, we denote it r-I-1.

Different $S_3$-symmetric, and softly broken, models allowing to accommodate DM were identified in Ref.~\cite{Khater:2021wcx}. Most of the models are ruled out due to unrealistic Yukawa sector. Possible DM candidates are (indicating an inert doublet and the Yukawa Lagrangian):
\begin{itemize}

\item R-I-1/r-I-1-$\mu_2^2$: $\mathrm{DM}\sim h_1$ or $\mathrm{DM}\sim (h_1,\,h_2)$, ${\mathcal{L}_Y \sim 1_f \otimes 1_h}$;

\item R-II-1a: $\mathrm{DM}\sim h_1$, ${\mathcal{L}_Y \sim 1_f \otimes 1_h}$~\cite{Khater:2021wcx};

\item r-III-s-$(\mu_2^2, \nu_{01}^2)$: $\mathrm{DM}\sim h_2$, ${\mathcal{L}_Y \sim (2 \oplus 1)_f \otimes (2 \oplus 1)_h}$ or ${\mathcal{L}_Y \sim 1_f \otimes 1_h}$;

\item C-III-a: $\mathrm{DM}\sim h_1$, ${\mathcal{L}_Y \sim 1_f \otimes 1_h}$;

\item c-III-b-$\mu_2^2$: $\mathrm{DM}\sim h_2$, ${\mathcal{L}_Y \sim (2 \oplus 1)_f \otimes (2 \oplus 1)_h}$ or ${\mathcal{L}_Y \sim 1_f \otimes 1_h}$;

\item c-III-c-$(\mu_2^2, \nu_{12}^2)$: $\mathrm{DM}\sim h_S$, ${\mathcal{L}_Y \sim (2 \oplus 1)_f \otimes 2_h}$;

\item c-IV-a-$(\mu_2^2, \nu_{01}^2)$: $\mathrm{DM}\sim h_2$, ${\mathcal{L}_Y \sim (2 \oplus 1)_f \otimes (2 \oplus 1)_h}$ or ${\mathcal{L}_Y \sim 1_f \otimes 1_h}$;

\end{itemize}

An R-I-1-like model was studied in Refs.~\cite{Machado:2012ed,Fortes:2014dca}. The vacuum of the model is given by $\left(0,\,0,\,w_S\right)$. In order to stabilise the $h_2$ doublet the authors imposed $\lambda_4=0$. Moreover, there are 3 pairs of mass-degenerate states, both neutral and charged, present between the $h_1$ and $h_2$ doublets. The degeneracy was lifted after introducing soft symmetry-breaking terms. It was found that this model may give rise to a viable DM candidate. 

The R-II-1a model was studied in Ref.~\cite{Khater:2021wcx}. The neutral scalar eigenstates of the inert doublet (DM candidate), $h_1$, correspond to mass eigenstates. There is no mixing between those states and they have opposite CP parities. Therefore, either of the particles could potentially be a DM candidate, whichever is lighter. The numerical analysis led to the conclusion that only one of these particles could be a good dark matter candidate. The one for which the mass is proportional to $\lambda_4$ was excluded. The range compatible with the applied constraints was identified to be $m_\mathrm{DM} \in [52.5,\,89]~\text{GeV}.$ Unlike the case for the IDM-like models, depicted in figure~\ref{Fig:mass-ranges}, where a viable DM high-mass region is present, this is not the case for R-II-1a. The main reason for this fact is that the inert-active scalar portal of R-II-1a is constrained by the underlying $S_3$ symmetry rendering it impossible to adjust it at higher DM masses.

In this work we shall consider the C-III-a model. In contrast to the aforementioned models the C-III-a vacuum allows for a nontrivial phase. This solution violates CP spontaneously~\cite{Emmanuel-Costa:2016vej}.

%%%%%%%%%%%%%%%%%%%%%%%%%%%%%%%%%%%%%%%%%%%%%%%%
\section{The C-III-a model}\label{sect:C-III-1a}
%%%%%%%%%%%%%%%%%%%%%%%%%%%%%%%%%%%%%%%%%%%%%%%%

%%%%%%%%%%%%%%%%%%%%%%%%%%%%%%%%%%%%%%%%%%%%%%%%
\subsection{Generalities}
%%%%%%%%%%%%%%%%%%%%%%%%%%%%%%%%%%%%%%%%%%%%%%%%

The C-III-a vacuum is defined by \cite{Emmanuel-Costa:2016vej}
\begin{equation}\label{CIIIaVEV}
\{w_1,\,w_2,\,w_S\}=
\{0,\,\hat{w}_2 e^{i \sigma},\,\hat{w}_S\},
\end{equation}
which is reminiscent of the R-II-1a vacuum, $\{0,\,w_2,\,w_S\} \in \mathbb{R}\mathrm{e}$. The only difference is that $w_2$ is complex. For complex cases ``hat'', $\hat w_i$, refers to the absolute value.

The minimisation conditions are:
\begin{subequations}\label{Eq: C-III-a-Min_Con}
\begin{align}
\mu_0^2 &= -\frac{1}{2}\lambda_b\hat{w}_2^2 - \lambda_8 \hat{w}_S^2,\\
\mu_1^2 &= - \left( \lambda_1 + \lambda_3\right)\hat{w}_2^2 - \frac{1}{2}\left( \lambda_b - 8 \cos^2 \sigma \lambda_7 \right)\hat{w}_S^2,\\
\lambda_4 &= \frac{4 \cos \sigma\hat{w}_S}{\hat{w}_2}\lambda_7.
\label{eq:c-iii-a-lam4-lam7}
\end{align}
\end{subequations}
with
\begin{equation}
\lambda_b=\lambda_5+\lambda_6-2\lambda_7.
\end{equation}

The DM candidate resides in the $h_1$ doublet. The $\mathbb{Z}_2$ symmetry is preserved for
\begin{equation}\label{eq.Z2h1}
h_1 \to - h_1,\text{ or else }\{h_2,\,h_S\} \to -\{h_2,\,h_S\}.
\end{equation}

It is convenient to redefine the decomposition \eqref{Eq:hi_hS} of $h_2$ by extracting an overall phase,
\begin{equation} \label{eq:phase}
h_2=e^{i\sigma}\left(
\begin{array}{c}h_2^{\prime+}\\ (\hat w_2+\eta_2^\prime+i \chi_2^\prime)/\sqrt{2}.
\end{array}\right)
\end{equation}
In the sequel we omit the primes on $h_2^+$, $\eta_2$ and $\chi_2$.

A trivial Yukawa sector is assumed, ${\mathcal{L}_Y \sim 1_f \otimes 1_h}$, and thus the $S_3$ singlet is solely responsible for masses of fermions. Making $w_S$ a reference point, we define:
\begin{equation}
\tan\beta = \frac{\hat w_2}{\hat w_S}.
\end{equation}
The vevs can be parameterised as:
\begin{equation}\label{eq.BetaAngleDefined}
\hat w_2 = v \sin \beta,\quad
\hat w_S = v \cos \beta, \quad \hat w_2^2 + \hat w_S^2 = v^2.
\end{equation}

With the following rotation:
\begin{equation}\label{eq.Rbeta}
\begin{aligned}
\mathcal{R}_\beta = \frac{1}{v} \begin{pmatrix}
v & 0 & 0 \\
0 & \hat w_2 & \hat w_S \\
0 & -\hat w_S & \hat w_2
\end{pmatrix} = & \begin{pmatrix}
1 & 0 & 0\\
0 & \sin \beta & \cos \beta \\
0 & -\cos \beta & \sin \beta
\end{pmatrix},
\end{aligned}
\end{equation}
we have
\begin{equation}
\mathcal{R}_\beta \begin{pmatrix}
0 \\
\hat w_2  \\
\hat w_S 
\end{pmatrix} = \begin{pmatrix}
0 \\ v \\ 0
\end{pmatrix}.
\end{equation}

Compared with R-II-1a, this model has one more parameter. The C-III-a vacuum acquires a non-vanishing relative phase $\sigma$. This comes at the ``cost'' of an additional constraint among two quartic terms, eq.~(\ref{eq:c-iii-a-lam4-lam7}). In fact, if we use this constraint for $\cos \sigma=1$, the expressions for $\mu_0^2$ and $\mu_1^2$ coincide between R-II-1a and C-III-a. For convenience we list the R-II-1a minimisation conditions:
\begin{subequations}
\begin{align}
\text{R-II-1a:} \quad \mu _0^2&= \frac{1}{2}\lambda _4\frac{ w_2^3}{w_S}-\frac{1}{2} \lambda_a w_2^2-\lambda _8 w_S^2, \\
\text{R-II-1a:} \quad \mu _1^2&= -\left( \lambda _1+ \lambda _3\right) w_2^2+\frac{3}{2} \lambda _4 w_2 w_S-\frac{1}{2} \lambda_a w_S^2,
\end{align}
\end{subequations} with $\lambda_a=\lambda_5+\lambda_6+2\lambda_7$. However, there is a subtlety, discussed in section~\ref{Sec:C-III-a_Inert_Sector}, that forces $\sigma \neq 0$ for C-III-c. This special limit will be discussed in section~\ref{sect:R-II-1a-vs-C-iii-c}.

%%%%%%%%%%%%%%%%%%%%%%%%%%%%%%%%%%%%%%%%%%%%%%%%
\subsection{C-III-a masses}\label{Sec:C-III-a_masses}
%%%%%%%%%%%%%%%%%%%%%%%%%%%%%%%%%%%%%%%%%%%%%%%%

%%%%%%%%%%%%%%%%%%%%%%%%%%%%%%%%%%%%%%%%%%%%%%%%
\subsubsection{Charged mass-squared matrix}
%%%%%%%%%%%%%%%%%%%%%%%%%%%%%%%%%%%%%%%%%%%%%%%%
    
The charged mass-squared matrix in the $\{h_1^+,\,h_2^+,\,h_S^+ \}$ basis is given by:
\begin{equation}
\mathcal{M}^2_\mathrm{Ch}
=\begin{pmatrix}
\left(\mathcal{M}_\mathrm{Ch}^2\right)_{11} & 0 & 0 \\
0 & \left(\mathcal{M}_\mathrm{Ch}^2\right)_{22} & \left(\mathcal{M}_\mathrm{Ch}^2\right)_{23} \\
0 & \left(\mathcal{M}_\mathrm{Ch}^2\right)_{23} & \left(\mathcal{M}_\mathrm{Ch}^2\right)_{33}
\end{pmatrix},
\end{equation}
where
\begin{subequations}
\begin{align}
\left(\mathcal{M}_\mathrm{Ch}^2\right)_{11} &= -2\lambda_3\hat w_2^2 -\frac{1}{2}\left[\lambda_6 - 10\lambda_7 - 8 \lambda_7 \cos(2\sigma)\right]\hat w_S^2, \\
\left(\mathcal{M}_\mathrm{Ch}^2\right)_{22} &= -\frac{1}{2}(\lambda_6-2\lambda_7)\hat w_S^2 , \\
\left(\mathcal{M}_\mathrm{Ch}^2\right)_{23} &= \frac{1}{2}(\lambda_6-2\lambda_7)\hat w_2 \hat w_S, \\
\left(\mathcal{M}_\mathrm{Ch}^2\right)_{33} &= -\frac{1}{2}(\lambda_6-2\lambda_7)\hat w_2^2.
\end{align}
\end{subequations}
The charged mass-squared matrix is diagonalisable by eq.~\eqref{eq.Rbeta}. The physical scalar states are given by:
\begin{subequations}
\begin{align}
h^+ & = h_1^+,\\
G^+ & = \sin\beta\,h_2^+ + \cos\beta\,h_S^+,\\ 
H^+ & = -\cos\beta\,h_2^+ + \sin\beta\,h_S^+,
\end{align}
\end{subequations}
with masses:
\begin{subequations} \label{Eq:C-III-c-masses-ch}
\begin{align}
m^2_{h^+} &= -2\lambda_3\hat w_2^2 -\frac{1}{2}\left[\lambda_6 - 10\lambda_7 - 8 \lambda_7 \cos(2\sigma)\right]\hat w_S^2, \\
m^2_{H^+} &= -\frac{1}{2}(\lambda_6-2\lambda_7)v^2.
\end{align}
\end{subequations}
Positivity of the mass-squared parameters requires the following constraints to be satisfied:
\begin{subequations}
\begin{align}
\lambda_6 & < -4 \lambda_3 \tan^2 \beta + 2 \lambda_7 \left[ 5 + 4 \cos(2\sigma) \right],\\
\lambda_6 & < 2 \lambda_7.
\end{align}
\end{subequations}

%%%%%%%%%%%%%%%%%%%%%%%%%%%%%%%%%%%%%%%%%%%%%%%%
\subsubsection{Inert-sector neutral mass-squared matrix}\label{Sec:C-III-a_Inert_Sector}
%%%%%%%%%%%%%%%%%%%%%%%%%%%%%%%%%%%%%%%%%%%%%%%%

The inert sector mass-squared matrix is in the $\{\eta_1,\,\chi_1\}$ basis given by:
\begin{equation}
\mathcal{M}^2_\mathrm{N1}
=\begin{pmatrix}
\left(\mathcal{M}_\mathrm{N1}^2\right)_{11} & \left(\mathcal{M}_\mathrm{N1}^2\right)_{12} \\
\left(\mathcal{M}_\mathrm{N1}^2\right)_{12} & \left(\mathcal{M}_\mathrm{N1}^2\right)_{22}
\end{pmatrix},
\end{equation}
where
\begin{subequations}
\begin{align}
\left(\mathcal{M}_\mathrm{N1}^2\right)_{11} &= -2 \left( \lambda_2 + \lambda_3 \right)\sin^2\sigma \hat w_2^2 + 2 \lambda_7 \left[ 5 + 4 \cos(2\sigma) \right] \hat w_S^2, \\
\left(\mathcal{M}_\mathrm{N1}^2\right)_{12} &= \left[ \left( \lambda_2 + \lambda_3 \right)\hat w_2^2 + 2\lambda_7 \hat w_S^2 \right] \sin(2\sigma), \\
\left(\mathcal{M}_\mathrm{N1}^2\right)_{22} &= -2\left[\left( \lambda_2 + \lambda_3 \right)\hat w_2^2 - 4\lambda_7 \hat w_S^2 \right] \cos^2\sigma.
\end{align}
\end{subequations}

This mass-squared matrix is diagonalisable 
\begin{equation}
\mathcal{R}_\gamma \mathcal{M}^2_\mathrm{N1} \mathcal{R}_\gamma^\mathrm{T} = \hat{\mathcal{M}}^2_\mathrm{N1},
\end{equation}
by
\begin{equation}
\mathcal{R}_\gamma = \begin{pmatrix}
\cos \gamma & \sin \gamma \\
-\sin \gamma & \cos \gamma
\end{pmatrix},
\end{equation}
where
\begin{equation} \label{Eq:tan_2gamma}
\tan (2 \gamma) = \frac{\left[ \left( \lambda_2 + \lambda_3 \right)\hat w_2^2 + 2 \lambda_7 \hat w_S^2 \right]\sin(2\sigma)}{ \left( \lambda_2 + \lambda_3 \right)\cos(2\sigma) \hat w_2^2 + \lambda_7 \left[ 3 + 2 \cos(2\sigma) \right] \hat w_S^2}.
\end{equation}

The physical neutral states are: 
\begin{subequations}
\begin{align}
\varphi_1 & = \cos\gamma\,\eta_1 + \sin\gamma\,\chi_1,\\ 
\varphi_2 & = -\sin\gamma\,\eta_1+ \cos\gamma\,\chi_1,
\end{align}
\end{subequations}
with masses
\begin{subequations} \label{Eq:inert-masses}
\begin{align}
m^2_{\varphi_1}=&-2(\lambda_2+\lambda_3)\hat w_2^2\sin^2(\gamma-\sigma) \nonumber \\
&+\lambda_7\hat w_S^2[7+6\cos(2\sigma)+3\cos(2\gamma)+2\cos(2\gamma-2\sigma)], \\
m^2_{\varphi_2}=&-2(\lambda_2+\lambda_3)\hat w_2^2\cos^2(\gamma-\sigma) \nonumber \\
&+\lambda_7\hat w_S^2[7+6\cos(2\sigma)-3\cos(2\gamma)-2\cos(2\gamma-2\sigma)].
\label{Eq:m_varphi2}
\end{align}
\end{subequations}

Equations (\ref{Eq:C-III-c-masses-ch}) and (\ref{Eq:inert-masses}) allow us to express $\lambda_2$, $\lambda_3$, $\lambda_6$ and $\lambda_7$ in terms of the four squared masses $m^2_{h^+}$, $m^2_{H^+}$, $m^2_{\varphi_1}$ and $m^2_{\varphi_2}$, as will be done in appendix~\ref{App:C-III-a-lambdas}.
On the other hand, if one takes $\lambda$'s as input, one finds that,
\begin{equation}
m_{\varphi_i}^2  = -\left( \lambda_2 + \lambda_3 \right)\hat w_2^2  
+ \lambda_7 \left[ 7 + 6 \cos(2\sigma) \right]\hat w_S^2 \mp \Delta,
\end{equation}
where
\begin{equation} \label{Eq:C-III-a-inert-splitting}
\Delta^2 = \left[ \left( \lambda_2 + \lambda_3\right) \hat w_2^2 
+ \lambda_7 \left( 2 + 3 \cos(2\sigma) \right) \hat w_S^2\right]^2 + 9 \lambda_7 ^2 \sin^2 (2\sigma) \hat w_S^4. 
\end{equation}
To ensure positivity of $m_{\varphi_i}^2$, if not taken as an input, we need to impose a constraint on the $\lambda_7$ coupling. For $\cos\sigma\neq0$, we find
\begin{equation}
\lambda_7>\left(\lambda_2+\lambda_3\right)
\frac{\tan^2 \beta}{4 \cos^2 \sigma}.
\end{equation}

Substituting the results for $\lambda$'s from appendix~\ref{App:C-III-a-lambdas} into the expression (\ref{Eq:tan_2gamma}), we find
\begin{subequations}\label{eq:m_phi2_vs_m_phi1}
\begin{equation}
f_+(\sigma,\gamma) m^2_{\varphi_1} = f_-(\sigma,\gamma) m^2_{\varphi_2},
\end{equation}
with
\begin{equation}
f_\pm(\sigma, \gamma) = [3+2\cos(2\sigma)]\sin(2\gamma-2\sigma)+\sin(2\gamma)\pm\sin(2\sigma).
\end{equation}
\end{subequations}

\begin{figure}[htb]
\begin{center}
%\vspace*{-4mm}
\includegraphics[scale=0.32]{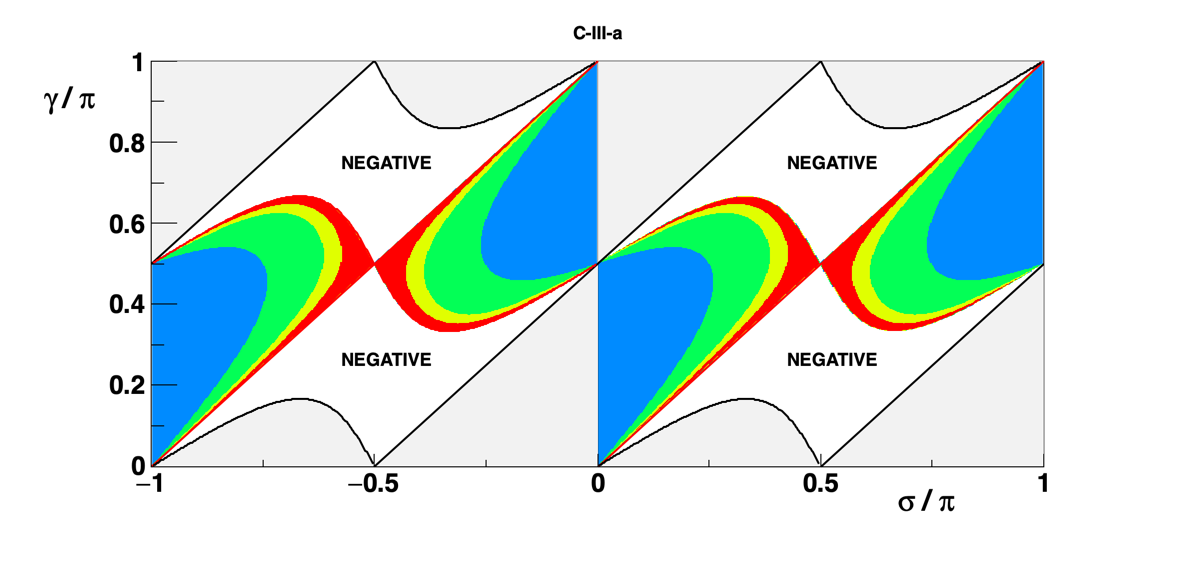}
\end{center}
\vspace*{-8mm}
\caption{The ratio $g(\gamma,\sigma)=m^2_{\varphi_2}/m^2_{\varphi_1}=f_+(\sigma,\gamma)/f_- (\sigma,\gamma)$ is shown for $g(\gamma,\sigma)>1$. Contours are shown at 2 (transition from blue to green), 5 (transition from green to yellow) and 10 (transition from yellow to red). According to eq.~(\ref{eq:m_phi2_vs_m_phi1}) the ratio depends on $\sigma$ only via the cosine and sine of $2\sigma$, and is thus the same for $\sigma$ and $\sigma+\pi$, as illustrated.}
\label{Fig:m2_vs_m1}
\end{figure}

In figure~\ref{Fig:m2_vs_m1} we show in colour regions where $m^2_{\varphi_2}>m^2_{\varphi_1}$. The red edge is where $m^2_{\varphi_1}/m^2_{\varphi_2}\to0$. In the white and grey regions, the ratio is either negative (white) or below 1 (grey). In fact, the latter region is identical to the coloured one, after a solid rotation by 180 degrees, $\{\sigma,\,\gamma\}\to\{\pi-\sigma,\,\pi/2-\gamma\}$, equivalent to an interchange of the two coefficients in eq.~(\ref{eq:m_phi2_vs_m_phi1}).

%%%%%%%%%%%%%%%%%%%%%%%%%%%%%%%%%%%%%%%%%%%%%%%%
%%%%%%%%%%%%%%%%%%%%%%%%%%%%%%%%%%%%%%%%%%%%%%%%

One observes from eq.~(\ref{Eq:C-III-a-inert-splitting}) that these states would become degenerate in the limit\footnote{Eq.~(\ref{eq:m_phi2_vs_m_phi1}) suggests that they might be near-degenerate in the limit $\sigma\to\epsilon$, with $\epsilon\ll1$. In this limit
\begin{subequations} \label{Eq:small-sigma}
\begin{align}
\lambda_2+\lambda_3&\simeq\frac{1}{12\hat w_2^2\sigma}[-(\sigma\cos2\gamma+2\sin2\gamma)(m^2_{\varphi_2}-m^2_{\varphi_1})
-\sigma(m^2_{\varphi_1}+m^2_{\varphi_2})], \\
\lambda_7&\simeq\frac{1}{24\hat w_S^2\sigma}[(\sigma\cos2\gamma-\sin2\gamma)(m^2_{\varphi_2}-m^2_{\varphi_1})
+\sigma(m^2_{\varphi_1}+m^2_{\varphi_2})],
\end{align}
\end{subequations}
so degeneracy actually requires $m^2_{\varphi_2}\to m^2_{\varphi_1}\to0.$}
\begin{subequations}
\begin{equation}
 \left( \lambda_2 + \lambda_3\right) \hat w_2^2 + \lambda_7 \left[ 2 + 3 \cos(2\sigma) \right] \hat w_S^2\to0,
\end{equation}
if simultaneously
\begin{equation}
\lambda_7 \sin(2\sigma) \hat w_S^2\to0.
\end{equation}
\end{subequations}
However, this limit is only reached for $\lambda_2+\lambda_3\to0$ and $\lambda_7\to0$, corresponding to massless states.

\begin{figure}[htb]
\begin{center}
%\vspace*{-4mm}
\includegraphics[scale=0.32]{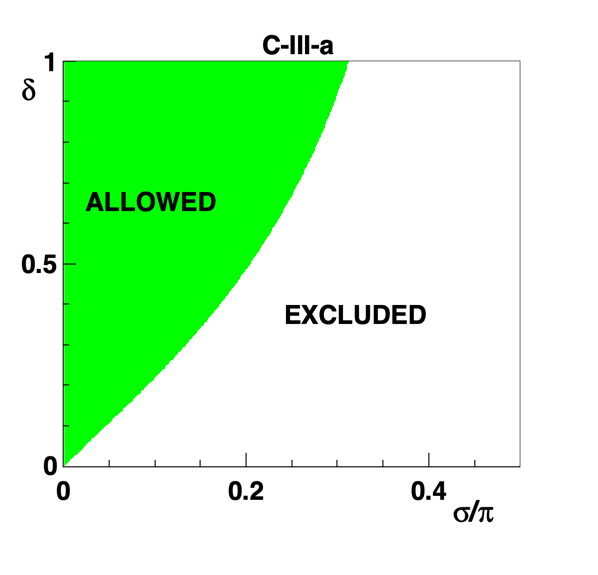}
\end{center}
\vspace*{-10mm}
\caption{The mass gap $\delta$ vs.\ $\sigma$ in the neutral inert sector. The green region is allowed.}
\label{Fig:massgap}
\end{figure}

%%%%%%%%%%%%%%%%%%%%%%%%%%%%%%%%%%%%%%%%%%%%%%%%
\paragraph{The mass gap.}
%%%%%%%%%%%%%%%%%%%%%%%%%%%%%%%%%%%%%%%%%%%%%%%%
Eliminating $\gamma$ from the equations (\ref{Eq:C-III-a-lambdas-a}), one finds expressions for $\lambda_2$, $\lambda_3$, $\lambda_6$ and $\lambda_7$ involving a square root, the argument of which must be positive:
\begin{equation}
9(m_{\varphi_2}^2-m_{\varphi_1}^2)^2-4m_{\varphi_1}^2m_{\varphi_2}^2\tan^2\sigma>0.
\end{equation}
For finite values of $\sigma$ this condition can be re-phrased as a condition on the mass gap 
\begin{equation}
\delta=\frac{m_{\varphi_2}^2-m_{\varphi_1}^2}{\sqrt{m_{\varphi_1}^2m_{\varphi_2}^2}}>\frac{2}{3}|\tan\sigma|,
\end{equation}
shown in figure~\ref{Fig:massgap}. Indeed, for a fixed value of $\sigma$ the absolute mass gap is proportional to the absolute mass scale. This poses a challenge for the high-mass region, see Fig.~\ref{Fig:mass-ranges}, where the electroweak precision data constrain the mass splitting.

%%%%%%%%%%%%%%%%%%%%%%%%%%%%%%%%%%%%%%%%%%%%%%%%
%%%%%%%%%%%%%%%%%%%%%%%%%%%%%%%%%%%%%%%%%%%%%%%%

%%%%%%%%%%%%%%%%%%%%%%%%%%%%%%%%%%%%%%%%%%%%%%%%
\subsubsection{Non-inert-sector neutral mass-squared matrix}
%%%%%%%%%%%%%%%%%%%%%%%%%%%%%%%%%%%%%%%%%%%%%%%%

The neutral mass-squared matrix in the basis of $\{\eta_2, \,\eta_S,\, \chi_2, \,\chi_S\}$ is given by: 
\begin{equation}
\mathcal{M}^2_\mathrm{N2S}
=\begin{pmatrix}
\left(\mathcal{M}^2_\mathrm{N2S}\right)_{11} & \left(\mathcal{M}^2_\mathrm{N2S}\right)_{12} & \left(\mathcal{M}^2_\mathrm{N2S}\right)_{13} & \left(\mathcal{M}^2_\mathrm{N2S}\right)_{14}\vspace{2pt} \\ 
\left(\mathcal{M}^2_\mathrm{N2S}\right)_{12} & \left(\mathcal{M}^2_\mathrm{N2S}\right)_{22} & \left(\mathcal{M}^2_\mathrm{N2S}\right)_{14} & \left(\mathcal{M}^2_\mathrm{N2S}\right)_{24}\vspace{2pt} \\ 
\left(\mathcal{M}^2_\mathrm{N2S}\right)_{13} & \left(\mathcal{M}^2_\mathrm{N2S}\right)_{14} & \left(\mathcal{M}^2_\mathrm{N2S}\right)_{33} & \left(\mathcal{M}^2_\mathrm{N2S}\right)_{34}\vspace{2pt} \\ 
\left(\mathcal{M}^2_\mathrm{N2S}\right)_{14} & \left(\mathcal{M}^2_\mathrm{N2S}\right)_{24} & \left(\mathcal{M}^2_\mathrm{N2S}\right)_{34} & \left(\mathcal{M}^2_\mathrm{N2S}\right)_{44}\vspace{2pt} \\ 
\end{pmatrix},
\end{equation}
where
\begin{subequations}
\begin{align}
\left(\mathcal{M}^2_\mathrm{N2S}\right)_{11} &= 2 \left( \lambda_1 + \lambda_3 \right)\hat w_2^2 - 6 \lambda_7 \cos^2 \sigma \hat w_S^2,\\
\left(\mathcal{M}^2_\mathrm{N2S}\right)_{12} &= \left( \lambda_b - 2 \lambda_7 \cos^2 \sigma \right) \hat w_2 \hat w_S,\\
\left(\mathcal{M}^2_\mathrm{N2S}\right)_{13} &= \lambda_7 \sin (2\sigma) \hat w_S^2,\\
\left(\mathcal{M}^2_\mathrm{N2S}\right)_{14} &= -\lambda_7 \sin (2\sigma) \hat w_2 \hat w_S,\\
\left(\mathcal{M}^2_\mathrm{N2S}\right)_{22} &= 2 \left( \lambda_7 \cos^2 \sigma \hat w_2^2 + \lambda_8 \hat w_S^2 \right),\\
\left(\mathcal{M}^2_\mathrm{N2S}\right)_{24} &= \lambda_7 \sin (2\sigma) \hat w_2^2,\\
\left(\mathcal{M}^2_\mathrm{N2S}\right)_{33} &= 2\lambda_7 \sin^2 \sigma \hat w_S^2,\\
\left(\mathcal{M}^2_\mathrm{N2S}\right)_{34} &= -2 \lambda_7 \sin^2 \sigma \hat w_2 \hat w_S,\\
\left(\mathcal{M}^2_\mathrm{N2S}\right)_{44} &= 2 \lambda_7 \sin^2 \sigma \hat w_2^2.
\end{align}
\end{subequations}

Due to CP non-conservation, the physical scalars will be combinations of all fields $\{\eta_2, \,\eta_S,\, \chi_2, \,\chi_S\}$. In order to identify physical states we start by rotating $\mathcal{M}^2_\mathrm{N2S}$,
\begin{equation}\label{Eq:C-III-a_NeutralActiveHBRot}
\begin{pmatrix}
\phi_1 \\
\phi_2 \\
G^0 \\
\phi_3
\end{pmatrix} = \mathcal{I}_2 \otimes \begin{pmatrix}
\sin \beta & \cos \beta \\
-\cos \beta & \sin \beta
\end{pmatrix} \begin{pmatrix}
\eta_2 \\
\eta_S \\
\chi_2 \\
\chi_S
\end{pmatrix}.
\end{equation}
Upon identifying the Goldstone boson, $G^0$, the remaining $3\times3$ mass-squared matrix in the $\phi_i$ basis becomes
\begin{equation} \label{Eq:M_sq_3by3}
\mathcal{M}^{2}_\mathrm{\phi}
=\begin{pmatrix}
\left( \mathcal{M}^{2}_\mathrm{\phi} \right)_{11} & \left( \mathcal{M}^{2}_\mathrm{\phi} \right)_{12} & 0 \\
\left( \mathcal{M}^{2}_\mathrm{\phi} \right)_{12} & \left( \mathcal{M}^{2}_\mathrm{\phi} \right)_{22} & \left( \mathcal{M}^{2}_\mathrm{\phi} \right)_{23} \\
0 & \left( \mathcal{M}^{2}_\mathrm{\phi} \right)_{23} & \left( \mathcal{M}^{2}_\mathrm{\phi} \right)_{33}
\end{pmatrix},
\end{equation}
where 
\begin{subequations}
\begin{align}
(\mathcal{M}^2_\mathrm{\phi})_{11} &= \frac{2}{v^2} \left[ \left( \lambda_1 + \lambda_3 \right)\hat{w}_2^4 + \left( \lambda_b - 4 \lambda_7 \cos^2 \sigma \right)\hat{w}_2^2\hat{w}_S^2 + \lambda_8 \hat{w}_S^4 \right] , \label{Eq:RII1a_Mphi11}\\
(\mathcal{M}^2_\mathrm{\phi})_{12} &= \frac{-1}{v^2} \left[ \left(2\lambda_1 + 2\lambda_3 - \lambda_b \right)\hat{w}_2^3 \hat{w}_S + \left( \lambda_b  - 8 \lambda_7 \cos^2 \sigma - 2 \lambda_8 \right)\hat{w}_2 \hat{w}_S^3 \right],\label{Eq:RII1a_Mphi12}\\
(\mathcal{M}^2_\mathrm{\phi})_{22} &= \frac{2}{v^2} \left[ \left( \lambda_1 + \lambda_3 - \lambda_b + 2 \lambda_7 \cos^2\sigma + \lambda_8 \right)\hat{w}_2^2 \hat{w}_S^2 + \lambda_7 \cos^2 \sigma \left( \hat{w}_2^4-3\hat{w}_S^4 \right) \right],\\
(\mathcal{M}^2_\mathrm{\phi})_{23} &= v^2 \lambda_7 \sin(2\sigma),\\
(\mathcal{M}^2_\mathrm{\phi})_{33} &= 2 v^2 \lambda_7 \sin^2 \sigma.
\end{align}
\end{subequations}
This matrix, $\mathcal{M}_\phi^2$, can be diagonalised in terms of the $\mathcal{R}^0$ rotation
\begin{equation}\label{Eq:C-III-a_NeutralActiveHBRotR0Diag}
\begin{pmatrix}
H_1 \\
H_2 \\
H_3
\end{pmatrix} = \mathcal{R}^0 \begin{pmatrix}
\phi_1 \\
\phi_2 \\
\phi_3
\end{pmatrix},
\end{equation}
with $\mathcal{R}^0$ parameterised as
\begin{equation}\label{Eq:RII1a_RotR0}
\mathcal{R}^0 \equiv \begin{pmatrix}
1 & 0 & 0 \\
0 & \cos {\theta_3} & \sin {\theta_3} \\
0 & -\sin {\theta_3} & \cos {\theta_3}
\end{pmatrix} \begin{pmatrix}
\cos {\theta_2} & 0 & \sin {\theta_2} \\
0 & 1 & 0 \\
-\sin {\theta_2} & 0 & \cos {\theta_2}
\end{pmatrix} \begin{pmatrix}
\cos {\theta_1} & \sin {\theta_1} & 0 \\
-\sin {\theta_1} & \cos {\theta_1} & 0 \\
0 & 0 & 1\\
\end{pmatrix},
\end{equation}
where we impose on the three neutral scalar states $H_i$ the convention $m_{H_i} \leq m_{H_{i+1}}$. 

With $\lambda$'s as input, one could proceed to perform diagonalisation of $\mathcal{M}_\phi^2$. In order to have more control over the physical aspects one would start with one or two masses as input, together with several angles of the mixing matrix, and then determine $\lambda$'s. Such approach is discussed in appendix~\ref{sect:lambdas_1_5_8}.

%%%%%%%%%%%%%%%%%%%%%%%%%%%%%%%%%%%%%%%%%%%%%%%%
\subsubsection{Mass eigenstates}
%%%%%%%%%%%%%%%%%%%%%%%%%%%%%%%%%%%%%%%%%%%%%%%%

The SU(2) doublets in terms of the mass eigenstates are:
\begin{subequations}\label{Eq:C-III-a_Expanded_Mass_Eigenstates}
\begin{align}
h_1 &= e^{i \gamma}\begin{pmatrix}
h^+ \\
\left(\varphi_1 + i \varphi_2\right)/\sqrt{2}
\end{pmatrix},\\
\begin{split}
h_2 & = e^{i \sigma}\begin{pmatrix}
\sin \beta\,G^+ - \cos \beta\,H^+\\
\left( \sin \beta\, v  + i \sin \beta\, G^0 + \sum_{i=1}^3 \left[ \sin \beta\,\mathcal{R}^0_{i1} - \cos \beta\, \left( \mathcal{R}^0_{i2} + i \mathcal{R}^0_{i3} \right)\right] H_i  \right) /\sqrt{2}
\end{pmatrix}\\
& = e^{i \sigma}\begin{pmatrix}
\sin \beta\,G^+ - \cos \beta\,H^+\\
\left( \sin \beta\, v  + i \sin \beta\, G^0 + \sum_{i=1}^3 A_{2i} H_i \right)/\sqrt{2}
\end{pmatrix},
\end{split}\\
\begin{split}
h_S & = \begin{pmatrix}
\cos \beta\, G^+ + \sin \beta\, H^+\\
 \left( \cos \beta\, v + i \cos \beta\,G^0 + \sum_{i=1}^3 \left[ \cos \beta\,\mathcal{R}^0_{i1} + \sin \beta\,\left( \mathcal{R}^0_{i2} + i \mathcal{R}^0_{i3} \right) \right] H_i \right)/\sqrt{2}
\end{pmatrix}\\
& = \begin{pmatrix}
\cos \beta\, G^+ + \sin \beta\, H^+\\
\left( \cos \beta\, v + i \cos \beta\,G^0 + \sum_{i=1}^3 A_{Si} H_i \right)/\sqrt{2}
\end{pmatrix},
\end{split}\label{Eq:h_S-expansion}
\end{align}
\end{subequations} 
where $A_{ij}$ is a complex quantity, implicitly defined by these equations. For simplicity, we extracted the $\gamma$ phase from $h_1$. This lets $\varphi_1$ and $\varphi_2$ be interpreted as mass eigenstates.

%%%%%%%%%%%%%%%%%%%%%%%%%%%%%%%%%%%%%%%%%%%%%%%%
\subsection{The C-III-a couplings}\label{Sect:C-III-1a-couplings}
%%%%%%%%%%%%%%%%%%%%%%%%%%%%%%%%%%%%%%%%%%%%%%%%

Below, we quote the gauge and Yukawa couplings of the C-III-a model. The scalar-sector couplings are collected in appendix~\ref{App:Scalar_Couplings_CIIIa}.

%%%%%%%%%%%%%%%%%%%%%%%%%%%%%%%%%%%%%%%%%%%%%%%%
\subsubsection{Gauge couplings}
%%%%%%%%%%%%%%%%%%%%%%%%%%%%%%%%%%%%%%%%%%%%%%%%

The gauge-scalar interactions of the C-III-a model are:
\begin{subequations}
\begin{align}
\begin{split}
\mathcal{L}_{VVH} =& \left[ \frac{g}{2 \mathrm{c}_w}m_ZZ_\mu Z^\mu + g m_W W_\mu^+ W^{\mu-} \right] \sum_{i=1}^3 \mathcal{R}_{i1}^0 H_i ,\label{Eq:RII1a_VVH}
\end{split}\\
\begin{split}
\mathcal{L}_{VHH} =&-\frac{ g}{2 \mathrm{c}_w}Z^\mu \left( \sum_{i<j=2}^3 \left( \mathcal{R}_{i2}^0\mathcal{R}_{j3}^0-\mathcal{R}_{i3}^0\mathcal{R}_{j2}^0 \right) H_i  \overset\leftrightarrow{\partial_\mu} H_j 
+ \varphi_1  \overset\leftrightarrow{\partial_\mu} \varphi_2  \right)\\
& - \frac{g}{2}\bigg\{ i W_\mu^+ \left(\sum_{i=1}^3  \left( \mathcal{R}_{i2}^0 + i \mathcal{R}_{i3}^0 \right) H^- \overset\leftrightarrow{\partial^\mu} H_i 
+ h^- \overset\leftrightarrow{\partial^\mu} 
(\varphi_1 + i \varphi_2) \right) + \mathrm{h.c.} \bigg\}\\
& + \left[ i e A^\mu + \frac{i g}{2} \frac{\mathrm{c}_{2w}}{\mathrm{c}_w} Z^\mu \right] \left( H^+ \overset\leftrightarrow{\partial_\mu} H^- + h^+ \overset\leftrightarrow{\partial_\mu} h^- \right),\raisetag{101pt}
\end{split}\\
\begin{split}
\mathcal{L}_{VVHH} =& \left[ \frac{g^2}{8 \mathrm{c}_w^2}Z_\mu Z^\mu + \frac{g^2}{4} W_\mu^+ W^{\mu-} \right] \left( H_1^2 + H_2^2 + H_3^2 + \varphi_1^2 + \varphi_2^2\right)\\
& + \bigg\{ \left[ \frac{e g}{2} A^\mu W_\mu^+ - \frac{g^2}{2} \frac{\mathrm{s}_w^2}{\mathrm{c}_w}Z^\mu W_\mu^+ \right] \bigg( \sum_{i=1}^3  \left( \mathcal{R}_{i2}^0 + i \mathcal{R}_{i3}^0 \right) H_i H^- \\ 
& \hspace{175pt} + (\varphi_1 + i \varphi_2) h^-\bigg) + \mathrm{h.c.} \bigg\}\\
&+ \left[ e^2 A_\mu A^\mu + e g \frac{\mathrm{c}_{2w}}{\mathrm{c}_w}A_\mu Z^\mu + \frac{g^2}{4} \frac{\mathrm{c}_{2w}^2}{\mathrm{c}_w^2}Z_\mu Z^\mu + \frac{g^2}{2} W_\mu^- W^{\mu +} \right]\left(H^-H^+ + h^-h^+ \right).\raisetag{110pt}\label{Eq:CIIIa_SSVV}
\end{split}
\end{align}
\end{subequations}

In terms of the mass eigenstates~\eqref{Eq:C-III-a_NeutralActiveHBRotR0Diag}, the SM-like Higgs boson could be identified with one of the $H_i$ fields if $H_i$ happens to be the only field that couples to the gauge bosons in eq.~\eqref{Eq:RII1a_VVH}. Therefore, for a given $H_i$ to be the SM-like Higgs field, this would require
\begin{equation}\label{Eq:C-III-a_SM_lim_GS}
\mathcal{R}_{i1}^0  \to 1,
\end{equation}
where the rotation matrix $\mathcal{R}^0$ is orthogonal, and hence $(\mathcal{R}^0_{i1})^2+(\mathcal{R}^0_{i2})^2+(\mathcal{R}^0_{i3})^2=1$. This means that all other entries of the row $i$ and column $1$ of the matrix $\mathcal{R}^0$ in eq.~\eqref{Eq:RII1a_RotR0} would have to be zero. 

From eq.~\eqref{Eq:C-III-a_NeutralActiveHBRot} we may conclude that $\phi_1$ can be identified with the SM-like Higgs boson provided that it is already a mass eigenstate. The rotation given by eq.~\eqref{Eq:C-III-a_NeutralActiveHBRot} guarantees that it is $\phi_1$ together with $G^0$ that appear in the new basis as the neutral fields of the only doublet that acquires a vev. The field $\phi_1$ would be a physical field when $\left( \mathcal{M}_{\phi}^2 \right)_{12}$ of eq.~\eqref{Eq:RII1a_Mphi12} is zero and, as a result, its mass is then given by $\left( \mathcal{M}_{\phi}^2 \right)_{11}$ in eq.~\eqref{Eq:RII1a_Mphi11}. Imposing $\mathcal{R}_{i1}^0=1$ for any $i$ always leads to $H_i \equiv \phi_1$.

%%%%%%%%%%%%%%%%%%%%%%%%%%%%%%%%%%%%%%%%%%%%%%%%
\subsubsection{Yukawa couplings}\label{Sec:Yukawa_Sector}
%%%%%%%%%%%%%%%%%%%%%%%%%%%%%%%%%%%%%%%%%%%%%%%%

There are two possibilities to construct the Yukawa Lagrangian:
\begin{equation*}
\begin{aligned}
&\mathcal{L}_Y \sim (2 \oplus 1)_f \otimes (2 \oplus 1)_h, ~~ \text{and }\\
&\mathcal{L}_Y \sim 1_f \otimes 1_h.
\end{aligned}
\end{equation*}
Although the first option can give realistic fermion masses, the CKM matrix splits into a block-diagonal form. We consider the trivial representation for fermions\footnote{In our study neutrino masses are of no particular interest.}:
\begin{equation}\label{Eq:LYRII1a}
- \mathcal{L}_Y  =  \overline Q_{i\,L}^{\,0} y_{ij}^d h_S d_{j\,R}^{\,0} + \overline Q_{i\,L}^{\,0} y_{ij}^u \tilde{h}_S u_{j\,R}^{\,0} + \text{(leptonic sector)} + \mathrm{h.c.},
\end{equation}
where $\tilde{h}_S$ is the charge conjugated of $h_S$, i.e., ${\tilde{h}_S=i\sigma_2 h_S ^\ast}$. The superscript ``$0$'' on the fermion fields indicates weak-basis fields. 

For the trivial Yukawa sector, the CKM matrix, $V_\mathrm{CKM} = V_u^\dagger V_d$, can be easily fixed to match the experimental value. Moreover, there is natural flavour conservation since the symmetry, whenever the fermions are singlets of $S_3$, only allows for the fermions to couple to one of the scalar doublets. There are no tree-level flavour changing neutral currents. The scalar-fermion couplings can be extracted from eq.~\eqref{Eq:LYRII1a} by transforming into the fermion mass-eigenstate basis and multiplying the appropriate coefficients by $-i$:
\begin{subequations}\label{Eq:C-III-a-Yukawa}
\begin{align}
g \left( H_i \bar{u} u \right) &= \frac{m_u}{v} \left[ -i\left( \mathcal{R}_{i1}^0 + \mathcal{R}_{i2}^0 \tan \beta \right) - \gamma_5 \mathcal{R}_{i3}^0 \tan \beta \right],\\ 
g \left( H_i \bar{d} d \right) &= \frac{m_d}{v} \left[ -i\left( \mathcal{R}_{i1}^0 + \mathcal{R}_{i2}^0 \tan \beta \right) + \gamma_5 \mathcal{R}_{i3}^0 \tan \beta \right].
\end{align}
\end{subequations}
The leptonic Dirac mass terms lead to similar relations.

Due to the CP-indefinite nature of $H_i$, the scalar-fermion decay rate is given by
\begin{equation}\label{Eq:CIIIa_Yukawa_Hiff}
\begin{split}
\Gamma \left( H_i \to  \bar{f} f \right) = \frac{N_c m_{H_i} m_f^2}{8 \pi v^2} &\bigg[ \left( 1- 4 \frac{m_f^2}{m_{H_i}^2} \right)^{3/2} | \mathcal{R}_{i1}^0 + \mathcal{R}_{i2}^0 \tan \beta |^2\\ &~\,~+ \left( 1- 4 \frac{m_f^2}{m_{H_i}^2} \right)^{1/2} | \mathcal{R}_{i3}^0 \tan \beta |^2 \bigg],
\end{split}
\end{equation}
with $N_c$ the number of colours ($N_c=3$ for quarks and $N_c=1$ for leptons). We approximate the decay rate ratio of the SM-like Higgs boson to that of the SM as
\begin{equation}\label{Eq:CIIIa_Gamma_SFF_Ratio}
\kappa_{ff}^2 \approx | \mathcal{R}_{11}^0 + \mathcal{R}_{12}^0 \tan \beta |^2 + \left( 1- 4 \frac{m_f^2}{m_{h_\mathrm{SM}}^2} \right)^{-1} | \mathcal{R}_{13}^0 \tan \beta |^2.
\end{equation}
This equation will be used as a measure of the SM-like limit for the fermion couplings.

Finally, the charged scalar-fermion couplings are:
\begin{subequations}\label{Eq:R-CIIIa-Yukawa_Charged}
\begin{align}
g \left( H^+ \bar{u}_i d_j \right) & = i \frac{\sqrt{2}}{v} \tan \beta \left[ P_L m_u - P_R m_d  \right] \left( V_\mathrm{CKM} \right)_{ij},\\
g \left( H^- \bar{d}_i u_j \right) & = i \frac{\sqrt{2}}{v} \tan \beta \left[ P_R m_u - P_L m_d  \right] \left( V_\mathrm{CKM}^\dagger \right)_{ji},\\
g \left( H^+ \bar{\nu} l  \right) & = - i \frac{\sqrt{2} m_l}{v} \tan \beta P_R,\\
g \left( H^- \bar{l} \nu  \right) & = - i \frac{\sqrt{2} m_l}{v} \tan \beta P_L.
\end{align}
\end{subequations}
The structure of the charged scalar couplings resembles the 2HDM Type-I model, except that in our definition $\tan \beta$ is the inverse in the sense that the vev of the doublet that couples to the fermions appears in the denominator.

%%%%%%%%%%%%%%%%%%%%%%%%%%%%%%%%%%%%%%%%%%%%%%%%
\section{Relations among \boldmath$S_3$-based models}
\label{sect:R-II-1a-vs-C-iii-c}
%%%%%%%%%%%%%%%%%%%%%%%%%%%%%%%%%%%%%%%%%%%%%%%%
Some of the $S_3$-based models share certain properties, in particular C-III-a and R-II-1a, as will be discussed below.
%%%%%%%%%%%%%%%%%%%%%%%%%%%%%%%%%%%%%%%%%%%%%%%%
\subsection{Relation of the C-III-a model to other \boldmath$S_3$-based 3HDMs}
%%%%%%%%%%%%%%%%%%%%%%%%%%%%%%%%%%%%%%%%%%%%%%%%

The C-III-a model can be related to several other $S_3$-based models~\cite{Emmanuel-Costa:2016vej}, by considering special limits leading to models neither with $w_1$ proportional to $w_2$ nor with vanishing vevs $w_2$ or $w_S$. However, such relations cannot always be established. Further insights can be obtained by consulting Ref.~\cite{Emmanuel-Costa:2016vej}.

For $\cos\sigma=0$, the mass splitting between the neutral states of the C-III-a inert sector, eq.~(\ref{Eq:C-III-a-inert-splitting}), becomes
\begin{equation}
\Delta = |\left( \lambda_2 + \lambda_3 \right) \hat{w}_2^2 - \lambda_7 \hat{w}_S^2|,
\end{equation}
and one of the states of that sector becomes massless due to the O(2) symmetry originating from putting $\lambda_4=0$\cite{Kuncinas:2020wrn}, and definite CP parities. This case is equivalent to C-III-f $(\pm i \hat{w}_1, i \hat{w}_2, \hat{w}_S)$ or C-III-g $(\pm i \hat{w}_1, -i \hat{w}_2, \hat{w}_S)$, depending on the quadrant of the phase $\sigma$, with $\hat{w}_1 \ll v$. Then, for $\Delta=0$, and $\lambda_7 = \left( \lambda_2 + \lambda_3 \right) \tan^2 \beta$, both states become massless, irrespective of the value of $\lambda_7$. Due to an additional constraint in terms of $\lambda_7$, this configuration becomes equivalent to C-IV-b $( \hat{w}_1, \pm i \hat{w}_2, \hat{w}_S)$ with $\hat{w}_1 \ll v$. However, in the C-IV-b model only one massless state arises due to the O(2) symmetry~\cite{Kuncinas:2020wrn}. It should be noticed that one of the mass eigenvalues of C-IV-b explicitly depends on $\hat{w}_1^2$.

Some other vacua~\cite{Emmanuel-Costa:2016vej} of the form $(0,x,y)$ can be reached. The R-II-1a is a special case and is discussed in the following subsection. The only other real model with an equivalent vacuum is R-III $(w_1, w_2, w_S)$. It is impossible to reach this model as R-III would simultaneously require both $\sigma=0$ and $\lambda_4=0$. However, for this to be satisfied, the only possibility is to set $\lambda_7=0$, which is not required by R-III. Moving to the complex vacua, there are some other possible cases. The C-III-d $(\pm i \hat{w}_1, \hat{w}_2, \hat{w}_S)$ and C-III-e $(\pm i \hat{w}_1, -\hat{w}_2, \hat{w}_S)$ cases are not reachable as one of the minimisation constraints depends on the $\lambda_2+\lambda_3$ term, whereas C-III-a does not. Next, it is possible to reach \mbox{C-IV-d} $( \hat{w}_1 e^{i \sigma_1}, \pm \hat{w}_2 e^{i \sigma_1}, \hat{w}_S)$, which is real, by setting $\lambda_7=0$. In this case an additional O(2)$\otimes$U(1)$_{h_S}$ symmetry arises, see Ref.~\cite{Kuncinas:2020wrn}, which is spontaneously broken, yielding two massless states. Finally, when both $\lambda_2 + \lambda_3=0$ and $\lambda_7=0$ are satisfied, C-III-a becomes a special case of C-V $(\hat{w}_1 e^{i \sigma_1}, \hat{w}_2 e^{i \sigma_2}, \hat{w}_S)$, which is, actually, real. In this case there is an additional O(2)$\otimes$U(1)$_{h_1}\otimes$U(1)$_{h_2}\otimes$U(1)$_{h_S}$ symmetry.

An overview of the above relations is summarised in table~\ref{Table:CIIIaReducedComparison}.

%%%%%%%%%%%%%%%%%%%%%%%%%%%%%%%%%%%%%%%%%%%%%%
{{\renewcommand{\arraystretch}{1.3}
\begin{table}[b!]
\caption{Relations of the C-III-a model to other $S_3$-based models~\cite{Emmanuel-Costa:2016vej}. Most of the presented models, in the general form, do not require $w_1=0$, while C-III-a does. In light of this, models are treated in the special limit of $\hat w_1 \to 0$, along with the explicit (general) minimisation conditions. The O(2) symmetry arises when $\lambda_4=0$, in which case there is no spontaneous CP violation. Other continuous symmetries, if present, are specified. Massless states, in terms of a single scalar field~\eqref{Eq:hi_hS}, $m_{X_i}$, or in terms of the mixing of fields, $m_{X_i-X_j}$, are presented.}
\label{Table:CIIIaReducedComparison}
\begin{center}
\begin{tabular}{|c|c|l|}
\hline\hline
Model & Conditions & Comments \\ \hline \hline
\begin{tabular}[c]{@{}c@{}}R-II-1a\\ \footnotesize $(0, w_2, w_S )$\end{tabular} & \begin{tabular}[c]{@{}c@{}} $\sigma=0$ \end{tabular} & Special point in R-II-1a, $\lambda_4 = 4 \lambda_7 w_S/w_2$. \\ \hline
\begin{tabular}[c]{@{}c@{}}R-III\\ \footnotesize$(w_1, w_2, w_S )$ \end{tabular} & \begin{tabular}[c]{@{}c@{}}   \end{tabular}& \begin{tabular}[l]{@{}l@{}}Not reachable, $\lambda_7 \neq 0$ in R-III.\end{tabular} \\ \hline \hline
\begin{tabular}[c]{@{}c@{}}C-III-d,e\\ \footnotesize$(\pm i \hat{w}_1, \hat{w}_2, \hat{w}_S)$,\\ \footnotesize$(\pm i \hat{w}_1, -\hat{w}_2, \hat{w}_S)$ \end{tabular} & \begin{tabular}[c]{@{}c@{}} \end{tabular} & \begin{tabular}[l]{@{}l@{}}Not reachable. There are no\\ vanishing couplings in C-III-d,e.\end{tabular}\\ \hline
\begin{tabular}[c]{@{}c@{}}C-III-f,g\\ \footnotesize$(\pm i \hat{w}_1, i \hat{w}_2, \hat{w}_S)$,\\ \footnotesize$(\pm i \hat{w}_1, -i \hat{w}_2, \hat{w}_S)$ \end{tabular} & \begin{tabular}[c]{@{}c@{}} $\sigma = \pm \pi/2$,\\  $\lambda_4=0$\end{tabular} & \begin{tabular}[c]{@{}l@{}}Additional O(2) symmetry; $m_{\chi_{1}}=0$. \end{tabular} \\ \hline
\begin{tabular}[c]{@{}c@{}}C-IV-b\\ \footnotesize$( \hat{w}_1, \pm i \hat{w}_2, \hat{w}_S)$ \end{tabular} &\begin{tabular}[c]{@{}c@{}}$\sigma = \pm \pi/2,$\\  $\lambda_4=0$,\\ $\lambda_7 = \left( \lambda_2 + \lambda_3 \right) \hat{w}_2^2 / \hat{w}_S^2$ \end{tabular} & \begin{tabular}[l]{@{}l@{}} Exact C-IV-b: additional O(2) symmetry.\\ 
C-III-a limit: another massless state;\\ \hspace{68pt}  $m_{\eta_1}=m_{\chi_1}=0$. \end{tabular}  \\ \hline
\begin{tabular}[c]{@{}c@{}}C-IV-d\\ \footnotesize $( \hat{w}_1 e^{i \sigma_1}, \pm \hat{w}_2 e^{i \sigma_1}, \hat{w}_S)$ \end{tabular} & $\lambda_4=\lambda_7=0$ & \begin{tabular}[l]{@{}l@{}}  Additional O(2)$\otimes$U(1)$_{h_S}$ symmetry;\\ \qquad $m_{\eta_1-\chi_1}=m_{\chi_2-\chi_S}=0.$ \end{tabular} \\ \hline
\begin{tabular}[c]{@{}c@{}}C-V\\ \footnotesize$(\hat{w}_1 e^{i \sigma_1}, \hat{w}_2 e^{i \sigma_2}, \hat{w}_S)$ \end{tabular} & \begin{tabular}[c]{@{}c@{}} $\lambda_2+\lambda_3=\lambda_4=\lambda_7 = 0$ \end{tabular} & \begin{tabular}[l]{@{}l@{}}  Additional O(2)$\otimes$U(1)$_{h_1}\otimes$U(1)$_{h_2}\otimes$U(1)$_{h_S}$ \\symmetry; $m_{\eta_1}=m_{\chi_1}=m_{\chi_2-\chi_S}=0$. \end{tabular} \\ \hline \hline 
\end{tabular}
\end{center}
\end{table}}
%%%%%%%%%%%%%%%%%%%%%%%%%%%%%%%%%%%%%%%%%%%%%

%%%%%%%%%%%%%%%%%%%%%%%%%%%%%%%%%%%%%%%%%%%%%%%%
\subsection{R-II-1a vs C-III-a}
%%%%%%%%%%%%%%%%%%%%%%%%%%%%%%%%%%%%%%%%%%%%%%%%

Both R-II-1a and C-III-a have vevs of the form $(0,x,y)$:
\begin{equation*}
\text{R-II-1a: }(0, w_2, w_S ),\quad \text{C-III-a: }(0, \hat{w}_2 e^{i \sigma}, \hat{w}_S ).
\end{equation*}

In R-II-1a there is no mixing between $\eta_1$ and $\chi_1$, which are the neutral components of the $h_1$ doublet, and in addition the neutral mass squared matrix in the $\{h_2, h_S\}$ sector is $2 \times 2$ block diagonal in such a way that the CP-odd states do not mix with the CP even states. All physical neutral states in R-II-1a have definite CP parity. In the \mbox{C-III-a} vacuum there is no such separation and the physical neutral scalars are not CP eigenstates.

One might expect to recover all the R-II-1a  masses and mixing from those of C-III-a by
simply taking the limit $\sigma= 0$, but as can be seen from the results 
presented in the previous sections, this is not the case. One may wonder why the R-II-1a case is not trivially recovered from the C-III-a case by simply taking $\sigma$ equal to zero. The explanation is simple, one just has to look at the minimisation condition coming from the variation of $\sigma$ which requires: 
\begin{equation}
\hat{w}_2^2 \hat{w}_S \sin\sigma \left( \lambda_4 \hat{w}_2 - 4 \lambda_7 \hat{w}_S \cos \sigma \right) = 0.
\label{minsig}
\end{equation}

We have two factors and the minimisation conditions are satisfied either for $\sigma= 0$ leading to the real solution R-II-1a, or for $\lambda_4$ related to $\lambda_7$ by  eq.~(\ref{eq:c-iii-a-lam4-lam7}). There is no need to impose both conditions at the same time. R-II-Ia does not require this additional condition relating $\lambda_4$ to $\lambda_7$.

Imposing both $\sigma= 0$ and the condition given by eq.~(\ref{eq:c-iii-a-lam4-lam7}) at the same time would lead to physical states with definite CP parities. Furthermore, the CP-odd sector $(\chi_2,\chi_S)$ would become massless, i.e., an additional massless state would arise. The neutral sector of $h_1$ would also be diagonal.

%%%%%%%%%%%%%%%%%%%%%%%%%%%%%%%%%%%%%%%%%%%%%
\section{Model analysis}
\label{sect:exp-constraints}
%%%%%%%%%%%%%%%%%%%%%%%%%%%%%%%%%%%%%%%%%%%%%

The model is analysed using the following input:
\begin{itemize}
\item The lightest $H_i$ state is the SM-like Higgs with $m_{H_1} = 125.25$ GeV~\cite{Zyla:2020zbs};
\item The Higgs basis rotation angle $\beta \in [0,~\pi/2]$ and the phase $\sigma \in [-\pi,~\pi]$;
\item The diagonalisation angles $\gamma \in [0,~\pi]$, $\theta_2 \in [-\pi/2,~\pi/2]$, and $\theta_3 \in [-\pi/2,~\pi/2]$;
\item The charged scalar masses $m_{\varphi_i^\pm} \supset \{ m_{h^+},~m_{H^+}\} \in [0.07,~1]$ TeV;
\item The dark matter candidate $m_{\varphi_1} \in [0,~1]$ TeV;
\end{itemize}
We are not using all the mass parameters as input. The values of $\{m_{H_2},\,m_{H_3},\,m_{\varphi_2},\,\theta_1\}$ are calculated based on the input angles. By convention, the masses preserve the hierarchy based on indices. 

For the numerical parameter scan, both theoretical and experimental constraints are imposed. Based on the constraints, several cuts are defined and applied, in analogy with our companion paper~\cite{Khater:2021wcx}:
\begin{itemize}
\item Cut~1: perturbativity, stability, unitarity checks, LEP constraints;
\item Cut~2: SM-like gauge and Yukawa sector, electroweak precision observables and\\ $B$ physics;
\item Cut~3: $H_1 \to \{\mathrm{invisible},~\gamma \gamma \}$ decays, DM relic density, direct searches;
\end{itemize}
with each of the subsequent constraint being superimposed over the previous ones.

%%%%%%%%%%%%%%%%%%%%%%%%%%%%%%%%%%%%%%%%%%%%%%%%
\subsection{Cut 1 constraints}
%%%%%%%%%%%%%%%%%%%%%%%%%%%%%%%%%%%%%%%%%%%%%%%%

%%%%%%%%%%%%%%%%%%%%%%%%%%%%%%%%%%%%%%%%%%%%%%%%
\begin{figure}[htb]
\begin{center}
%\vspace*{-4mm}
\includegraphics[scale=0.3]{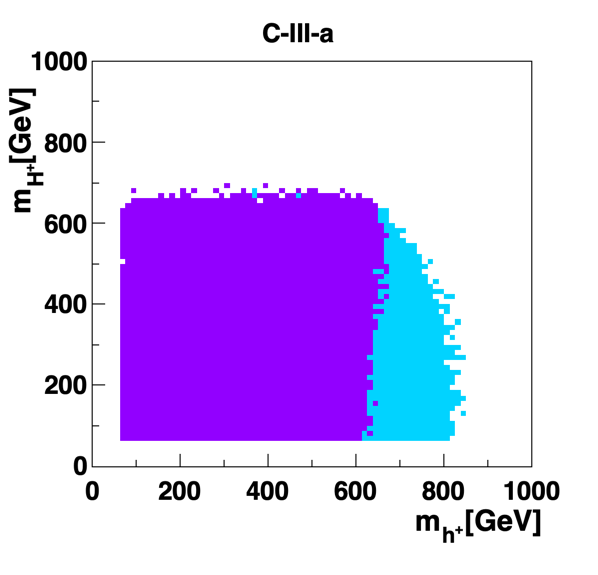}\\
\includegraphics[scale=0.3]{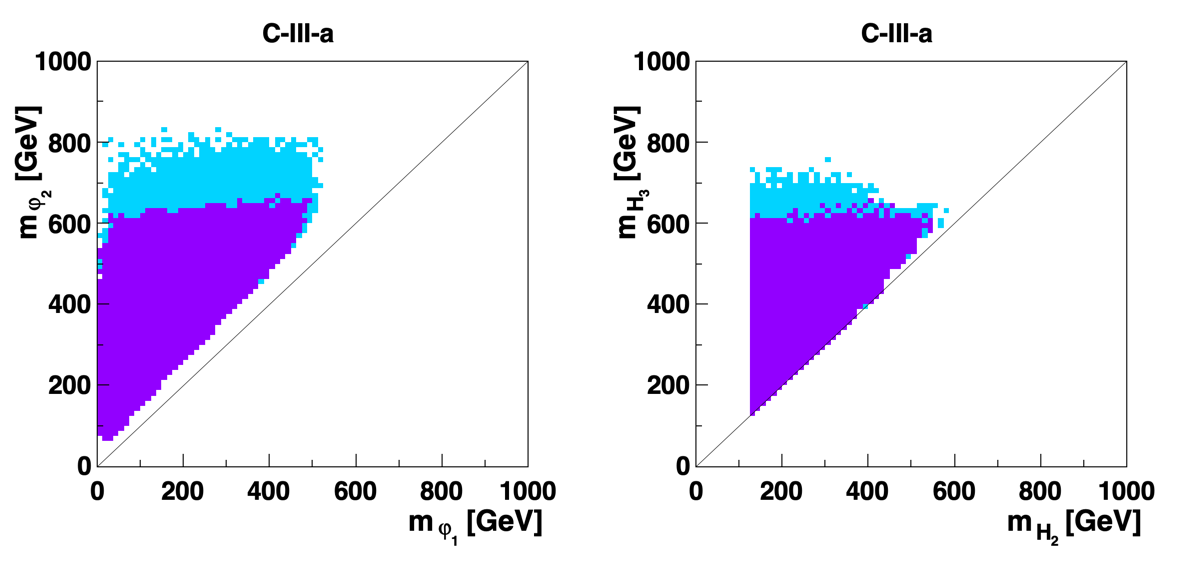}
\end{center}
\vspace*{-8mm}
\caption{Scatter plots of masses that satisfy the theory constraints, Cut~1.
Top: the charged sector, $h^\pm$ and $H^\pm$.
Bottom left: the inert neutral sector, $\varphi_1$ and $\varphi_2$.
Bottom right: the active heavy neutral sector, $H_2$ and $H_3$.
The light-blue region accommodates the $16\pi$ unitarity constraint, whereas the darker region satisfies the $8\pi$ constraint.}
\label{Fig:C-III-a-masses-th_constraints}
\end{figure}
%%%%%%%%%%%%%%%%%%%%%%%%%%%%%%%%%%%%%%%%%%%%%%%%

We start by putting constraints on the input masses. The mass of the SM-like Higgs particle is fixed at $m_{H_1} = 125.25$ GeV~\cite{Zyla:2020zbs}. In the extended Higgs sector studies a conservative lower bound for the charged masses is usually adopted as $m_{\varphi_i^\pm} \geq 80\text{ GeV}$~\cite{Pierce:2007ut,Arbey:2017gmh}. We shall assume a more generous value of $m_{\varphi_i^\pm} \geq 70\text{ GeV}$. Moreover, measurements of the $W^\pm$ and $Z$ widths at LEP~\cite{Schael:2013ita} forbid decays of the gauge bosons into a pair of scalars.  The lower limits on the scalar masses is set to be $m_{\varphi_i} + m_{h^\pm} > m_{W^\pm}$, and $m_{\varphi_1} + m_{\varphi_2} > m_Z$.

The theory constraints consist of several checks:
\begin{itemize}
\item \textbf{Unitarity} \\
The tree-level unitarity conditions for the $S_3$-symmetric 3HDM were presented in Ref.~\cite{Das:2014fea}. The unitarity limit is evaluated enforcing the absolute values of the eigenvalues $\Lambda_i$ of the scattering matrix  to be within a specific limit. In our scan we assume $|\Lambda_i| \leq 16\pi$~\cite{Lee:1977eg}. Some authors prefer a more severe bound $|\Lambda_i| \leq 8\pi$~\cite{Luscher:1988gc, Marciano:1989ns}. We compare the impact of both in figure~\ref{Fig:C-III-a-masses-th_constraints}.

\item \textbf{Perturbativity}\\
The perturbativity check is split into two parts: couplings are assumed to be within the limit $|\lambda_i| \leq 4\pi$ and the overall strength of the quartic scalar interactions is limited by $|g_{\varphi_i \varphi_j \varphi_k \varphi_l}| \leq4 \pi$. 

The list of quartic scalar interactions $g_{\varphi_i \varphi_j \varphi_k \varphi_l}$ can be found in appendix~\ref{App:Scalar_Couplings_CIIIa}. From the quartic interaction $h^\pm h^\pm h^\mp h^\mp$~\eqref{Eq:hphphmhm}, it follows that $0 < \lambda_1 + \lambda _3\leq \pi$. Evaluation of other couplings is more involved. An interesting observation, based on data satisfying Cut~1, is that $\lambda_7$ in \eqref{Eq:C-III-a-lambda7} must be poitive.

\item \textbf{Stability}\\
Necessary, but not sufficient, conditions for the stability of an $S_3$-symmetric 3HDM were provided in Ref.~\cite{Das:2014fea}. We parameterise the SU(2) doublets in terms of the spinor components,
\begin{equation}\label{Eq.StabilityGeneralDoublets}
h_i = || h_i || \hat{h}_i, \quad i = \{1, 2, S \},
\end{equation}
following the guideline presented in Refs.~\cite{ElKaffas:2006gdt,Grzadkowski:2009bt}. The complex product between two different unit spinors relies on six degrees of freedom. However, it was pointed out that those six variables are not independent, see section III-C of Ref.~\cite{Faro:2019vcd}. As a result, positivity conditions would yield an over-constrained $\lambda$ parameter space. In other words, the value of the potential would be lower than the true minimum due to additional parameters. To sum up, the norms of the spinors $|| h_i ||$ are parameterised in terms of the spherical coordinates
\begin{equation}
|| h_1 || = r \cos \gamma \sin \theta,  \qquad || h_2 || = r \sin \gamma \sin \theta, \qquad || h_S || = r \cos \theta,
\end{equation}
and the unit spinors are given by
\begin{equation}
\hat{h}_1 = \begin{pmatrix}
0 \\ 1
\end{pmatrix}, \qquad \hat{h}_2 = \begin{pmatrix}
\sin \alpha_2 \\ \cos \alpha_2 \, e^{i \beta_2}
\end{pmatrix},\qquad \hat{h}_S = e^{i \delta} 
\begin{pmatrix}
\sin \alpha_3 \\
\cos \alpha_3 \, e^{i \beta_3}.
\end{pmatrix},
\end{equation}

Due to the freedom of the $\lambda_4$ coupling the stability conditions are rather involved. Our approach involves checking the necessary stability constraints~\cite{Das:2014fea}, and if those are satisfied, with the help of the $\mathsf{Mathematica}$ function $\mathsf{NMinimize}$, using different algorithms, a further numerical minimisation of the potential is performed.

\end{itemize}

By imposing the theory constraints we exclude regions of the parameter space, as illustrated in figure~\ref{Fig:C-III-a-masses-th_constraints}.
Some masses, $m_{H^+}$, $m_{\varphi_1}$ and $m_{H_2}$, are cut off at high values by the perturbativity constraint, whereas $m_{h^+}$, $m_{\varphi_2}$ and $m_{H_3}$ are cut off by the unitarity constraint. As seen in the bottom-left panel of figure~\ref{Fig:C-III-a-masses-th_constraints}, a gap develops between the masses of the two neutral states of the inert sector, as discussed in section~\ref{Sec:C-III-a_Inert_Sector}. Experimental constraints will further reduce regions of the parameter space.

%%%%%%%%%%%%%%%%%%%%%%%%%%%%%%%%%%%%%%%%%%%%%%%%
\subsection{Cut 2 constraints}
%%%%%%%%%%%%%%%%%%%%%%%%%%%%%%%%%%%%%%%%%%%%%%%%

Cut~2 constraints are superimposed over those parameter points which pass the Cut~1 constraints. For a point to pass Cut~2, it needs to satisfy:
\begin{itemize}

\item \textbf{SM-like limit}

The SM-like limit for the gauge interactions was presented in eq.~\eqref{Eq:C-III-a_SM_lim_GS}, and the scalar-fermion decay rates were presented in eq.~\eqref{Eq:CIIIa_Yukawa_Hiff}. We recall that in the \mbox{C-III-a} model the active neutral scalars are CP-indefinite. In light of this, eq.~\eqref{Eq:CIIIa_Gamma_SFF_Ratio} is evaluated as a probe of the SM-like limit for Higgs-fermion couplings. We shall adopt the following 3-$\sigma$ bounds from the PDG~\cite{Zyla:2020zbs}:
\begin{subequations}\label{Eq:CIIIaSMLikeLimit}
\begin{align}
&\kappa^2_{VV} \equiv ({\cal R}^0_{11})^2\in\{1.19\pm3\,\sigma\},\text{ which comes from $h_\mathrm{SM} W^+ W^-$,}\\
&\kappa^2_{ff} \in\{1.04\pm3\,\sigma\},\text{ which comes from $h_\mathrm{SM} b\bar b$,}
\end{align}
\end{subequations}
where ${\cal R}^0_{11}=\cos\theta_1\cos\theta_2$. The gauge coupling depends only on two variables, which are $\theta_1$ and $\theta_2$. Nevertheless, there are other non-SM-like scalar gauge couplings present, which do not vanish, namely the trilinear $Z H_1 H_i$ and $W^\pm H^\mp H_1$, and quartic $\left(  A W^\pm + Z W^\pm\right) H^\mp H_1$. However, due to kinematics those do not contribute to the width of $H_1$. On the other hand, $\kappa_{ff}^2$ depends on $\{\theta_1,\, \theta_2, \, \beta\}$.

The 3-$\sigma$ allowed regions in  $\theta_1$, $\theta_2$ and $\beta$ are given in figure~\ref{Fig:SM-Like_Limit_C-III-a}. The angles $\theta_1$ and $\theta_2$ surviving Cut~2 tend to be small, whereas $\beta$ populates regions around $0.2\pi$ and $0.4\pi$. In our analysis values of $\theta_1$ are calculated while angles $\beta$ and $\theta_2$ are used as input.

%%%%%%%%%%%%%%%%%%%%%%%%%%%%%%%%%%%%%%%%%%%%%%%%
\begin{figure}[h]
\begin{center}
%\vspace*{-4mm}
\includegraphics[scale=0.35]{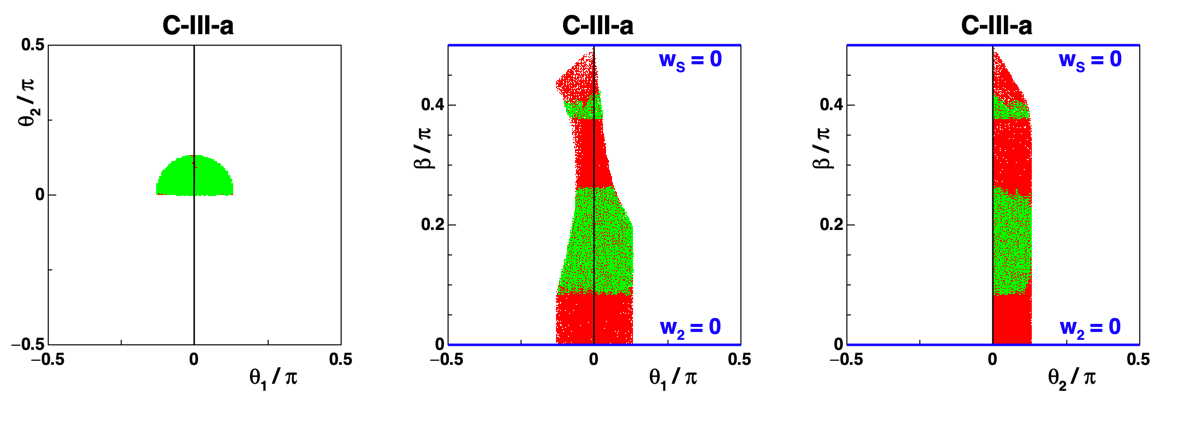}
\end{center}
\vspace*{-4mm}
\caption{Constraints on $\theta_1$, $\theta_2$ and $\beta$ from the gauge and Yukawa couplings. Red: values satisfying simultaneously $\kappa^2_{VV}$ and $\kappa^2_{ff}$ at 3-$\sigma$. Green: values satisfying Cut~2, plotted over the red-coloured background. Values of $\beta$ for which $\hat w_2$ or $\hat w_S$ vanish are identified by blue lines. The range of $\theta_2$ has been reduced due to the symmetry under $\theta_2 \to - \theta_2$.}
\label{Fig:SM-Like_Limit_C-III-a}
\end{figure}
%%%%%%%%%%%%%%%%%%%%%%%%%%%%%%%%%%%%%%%%%%%%%%%%

\item \textbf{Electroweak precision observables}

The electroweak oblique parameters are specified by the $S$, $T$, and $U$ functions~\cite{Peskin:1990zt,Peskin:1991sw}. Sufficient mass splittings of the extended electroweak sector can lead to a non-negligible contribution. The $S$ and $T$ parameters get the most sizeable contributions. Results are compared against the experimental constraints provided by the PDG~\cite{Zyla:2020zbs}, assuming that $U=0$. The model-dependent rotation matrices, needed to evaluate the set of $S$ and $T$, are presented in appendix~\ref{App:Peskin_Takeuchi_Rot}.  

\item \textbf{\boldmath$B$ physics constraints}

The importance of a charged scalar exchange for the $\bar B\to X(s)\gamma$ rate has been known since the late 1980's~\cite{Grinstein:1987pu,Hou:1988gv,Grinstein:1990tj}. Although three-Higgs-doublet models have two charged Higgs bosons, in the $S_3$-based models we are considering, only one of them couples to fermions, the other one is in the inert sector. This implies that we may follow the approach of Misiak and Steinhauser~\cite{Misiak:2006ab}, used for the 2HDM with relative Yukawa couplings of the active charged scalar, eq.~\eqref{Eq:R-CIIIa-Yukawa_Charged}, which in the notation of Ref.~\cite{Misiak:2006ab} corresponds to
\begin{equation} \label{Eq:Au-Ad}
A_u=A_d=\tan\beta,
\end{equation}
since, as pointed out before, the $\tan\beta$ here is the inverse of their $\tan\beta$.
According to eq.~(\ref{Eq:Au-Ad}) the relevant couplings are the same as those of the 2HDM Type~I model, with the exception that here we are interested in small values of $\tan\beta$. The $\bar B\to X(s)\gamma$ constraint excludes values of $|\tan\beta|$ larger than four. After applying Cut~3 the allowed range is shrunk to  $|\tan\beta|<1$.

We adopt techniques presented in the companion paper~\cite{Khater:2021wcx}. The experimental value is taken to be $\mathrm{Br} \left( \bar B\to X(s)\gamma  \right) \times 10^4 = 3.32 \pm 0.15$~\cite{Zyla:2020zbs}. We impose an $(n=3)$-$\sigma$ tolerance, together with an additional ten per cent computational uncertainty,
\begin{equation}
\begin{aligned}
\mathrm{Br} \left( \bar B\to X(s)\gamma  \right) \times 10^4 &=  3.32 \pm \sqrt{(3.32 \times 0.1)^2 + (0.15\,n)^2}\,.
\end{aligned}
\end{equation}
The acceptable region, corresponding to the 3-$\sigma$ bound, is $[2.76;\,3.88 ]$.
\end{itemize}

%%%%%%%%%%%%%%%%%%%%%%%%%%%%%%%%%%%%%%%%%%%%%%%%
\begin{figure}[htb]
\begin{center}
%\vspace*{-4mm}
\includegraphics[scale=0.3]{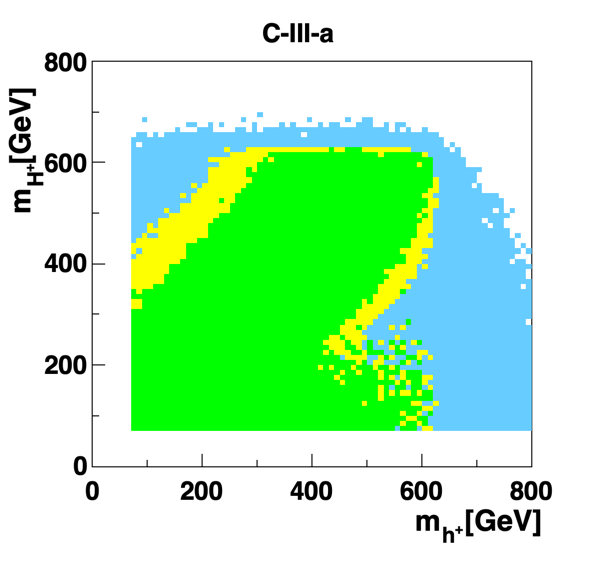}
\includegraphics[scale=0.3]{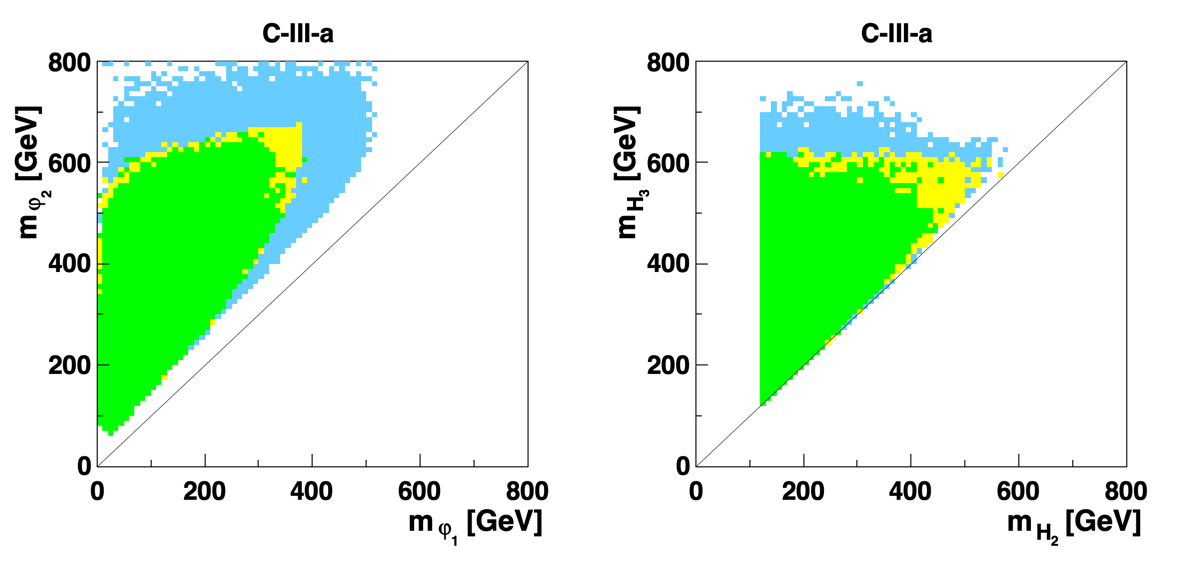}
\end{center}
\vspace*{-8mm}
\caption{Scatter plots of masses that satisfy Cut~1 and Cut~2 constraints.
Top: the charged sector, $h^\pm$ and $H^\pm$.
Bottom left: the inert neutral sector, $\varphi_1$ and $\varphi_2$.
Bottom right: the active heavy neutral sector, $H_2$ and $H_3$.
The light-blue region satisfies Cut~1 and accommodates the $16\pi$ unitarity constraint.
The yellow region accommodates a 3-$\sigma$ tolerance with respect to Cut~2,  whereas in the green regions, the model is within the 2-$\sigma$ bound of these values.}
\label{Fig:C-III-Cut2}
\end{figure}
%%%%%%%%%%%%%%%%%%%%%%%%%%%%%%%%%%%%%%%%%%%%%%%%

After applying Cut~2 the mass ranges of figure~\ref{Fig:C-III-a-masses-th_constraints} are reduced. The mass scatter plots satisfying Cut~1 and Cut~2 are presented in figure~\ref{Fig:C-III-Cut2}. The most obvious reduction of the allowed parameters is in the charged sector. The $\bar B\to X(s)\gamma $ constraint introduces cuts in two regions of the charged-Higgs masses, i.e., to the left and to the right of the allowed 3-$\sigma$ yellow, region. However, for relatively light charged scalars $H^\pm$ with $m_{H^\pm} \lesssim 300~\text{GeV}$, heavier $h^\pm$ states with $m_{h^\pm} > 600~\text{GeV}$, are allowed by the $\bar B\to X(s)\gamma $ constraint. However, this region, for heavy $h^\pm$ states, is excluded by the SM-like constraints and electroweak precision observables. On the other hand, heavy $H^{\pm}$ scalars are disfavoured by the electroweak precision observables. The upper-right corner of the $m_{\varphi_1}$-$m_{\varphi_2}$ Cut~1 plane (see figure~\ref{Fig:C-III-Cut2}) is excluded due to the $\bar B\to X(s)\gamma $ constraint. This is rather unexpected, since the constraint on the charged scalars would normally (in the IDM) not limit the parameter space of the neutral scalar sector. It arises due to the fact that the model parameters are highly constrained. Other regions of the Cut~1 mass scatter plot are excluded due to a combination of several Cut~2 constraints. Concerning the heavy active neutral sector, we note that the Cut~2 puts bounds on the upper value of the mass of $H_3$. 

%%%%%%%%%%%%%%%%%%%%%%%%%%%%%%%%%%%%%%%%%%%%%%%%
\subsection{Cut 3 constraints}
%%%%%%%%%%%%%%%%%%%%%%%%%%%%%%%%%%%%%%%%%%%%%%%%
This subsection includes constraints coming from the LHC and astrophysical observables. In the future, Higgs self interactions may become a crucial test, those are also discussed.
%%%%%%%%%%%%%%%%%%%%%%%%%%%%%%%%%%%%%%%%%%%%%%%%
\subsubsection{LHC Higgs constraints}
%%%%%%%%%%%%%%%%%%%%%%%%%%%%%%%%%%%%%%%%%%%%%%%%

First of all, we require that the full width of the SM-like Higgs particle be within $\Gamma_{H_1} = 3.2^{+2.8}_{-2.2}$~MeV, an experimental bound taken from the PDG \cite{Zyla:2020zbs}. In the SM the total width of the Higgs boson is around 4 MeV. The upper value, i.e., $\Gamma_{H_1} = 6~\text{MeV}$ is used in preliminary checks within the spectrum generator. Apart from that, several channels are checked against the experimental results:
\begin{itemize}
\item \textbf{Decay \boldmath$H_1 \to g g $ }

In the SM case, the dominant Higgs production mechanism is through gluon fusion. However, due to experimental limitations we do not explicitly consider constraints on this channel. For DM mass below $m_{H_1}/2$ the gluon branching ratio can become low due to the opening of the invisible channel, $H_1 \to \varphi_i\varphi_i$. However, such cases are partially excluded by other LHC Higgs-particle constraints of Cut~3. After applying all of the constraints we found that $\mathrm{Br}(H_1 \to gg) \in [6.5;\,8.4]\times 10^{-2}$, while the SM case predicts the value of $\mathrm{Br}(h^\mathrm{SM} \to gg) \approx 7.9 \times 10^{-2}$.

\item \textbf{Decay \boldmath$H_1 \to \gamma \gamma$}

The di-photon partial decay width is modified by the contributions of the charged-scalar loops which are not present in the SM. In light of the above discussion regarding gluons, we do not aim to account for the correct $H_1$ two-gluon production factor, instead we approximate the di-photon channel strength by
\begin{equation}\label{Eq:Diphoton_strength}
\mu_{\gamma\gamma} \approx \frac{\Gamma\left( H_1 \to \gamma \gamma \right)}{\Gamma^\text{exp}\left( h \to \gamma \gamma \right)} \frac{\Gamma^\text{exp} \left( h \right)}{\Gamma\left(H_1\right)},
\end{equation}
with $\mu_{\gamma \gamma} = 1.11 \pm 0.10$~\cite{Zyla:2020zbs}. We evaluate this constraint allowing for an additional ten per cent computational uncertainty, and impose an $(n=3)\text{-}\sigma$ tolerance,
\begin{equation}\label{Eq:Diphoton_Bounds}
\mu_{\gamma\gamma} = 1.11 \pm \sqrt{(1.11\times0.1)^2 + (0.1n)^2},
\end{equation}
which corresponds to the 3-$\sigma$ range of $[0.79;\,1.43]$. 

The di-photon branching ratio is higher for light $h^\pm$. As the mass of the inert sector charged scalar increases, the branching ratio also decreases. On the other hand, the di-photon branching ratio increases for heavier $H^\pm$ scalars.

\item \textbf{Invisible decays, \boldmath $H_1 \to\mathrm{inv.}$}

The SM-like Higgs boson can decay to lighter scalars, $H_1 \to \varphi_i \varphi_j$. If such decays are kinematically allowed, these processes can enhance the total width of the SM-like Higgs state sizeably. In total, due to CP non-conservation, and if kinematically accessible, there are three possible decay channels $H_1 \to \{\varphi_1 \varphi_1,\,\varphi_2 \varphi_2,\,\varphi_1 \varphi_2 \}$. After applying all of the cuts we found that \mbox{$m_{\varphi_2}\geq 150~\text{GeV}$}. This lower mass limit significantly simplifies the study of invisible decay channels, since the only accesible channel will be $H_1 \to \varphi_1 \varphi_1$. Furthermore, after applying all constraints, including Cut~3, the \mbox{$m_{\varphi_1}< m_{H_1}/2$} inequality always holds. Hence, the invisible decays channel is always open in the C-III-a model. In our calculations we adopt the PDG~\cite{Zyla:2020zbs} constraint, which is $\text{Br}^\mathrm{exp}\left( H_1 \to \text{inv.} \right)<0.19$.

We note that there are more severe constraints on the invisible channel than those appearing in the PDG, set by ATLAS~\cite{ATLAS:2020cjb,ATLAS:2020kdi}, $\text{Br}\left( H_1 \to \text{inv.} \right) \lesssim 0.13$. However, those are preliminary results. Even after applying a more strict bound we found a very limited impact on the parameter space. 

\end{itemize}

Analytic expressions for the decay rates presented in this section can be found in Appendix~\ref{App:DiGamma}.

%%%%%%%%%%%%%%%%%%%%%%%%%%%%%%%%%%%%%%%%%%%%%%%%
\subsubsection{ The \boldmath$H_1$ scalar self interactions}
%%%%%%%%%%%%%%%%%%%%%%%%%%%%%%%%%%%%%%%%%%%%%%%%

Let us next consider the trilinear and quadrilinear self interactions of the SM-like Higgs particle. In the future, the trilinear interactions may become a crucial test for new physics \cite{Bahl:2022jnx}. In the SM the Higgs self-interactions are~\cite{Boudjema:1995cb}:
\begin{equation}
g(h^3_\mathrm{SM})=\frac{3m_{h_\mathrm{SM}} ^2}{v},\quad
g(h_\mathrm{SM}^4)
= \frac{1}{v}g(h_\mathrm{SM}^3).
\end{equation}
In the C-III-a model the corresponding couplings are given by eqs.~\eqref{Eq.C_III_a_HiHjHk} and \eqref{Eq.C_III_a_HiHjHkHl}. Invoking the expressions for the $\lambda$'s given in appendix~\ref{App:C-III-a-lambdas}, as well as eq.~\eqref{eq:c-iii-a-lam4-lam7}, we find that the trilinear coupling can be expanded as
\begin{equation}
g(H_1^3) = \frac{1}{v} \left[ m_{H_1}^2 A_{H_1} + m_{H_2}^2 A_{H_2} + m_{H_3}^2 A_{H_3} + m_{\varphi_1}^2 A_{\varphi_1}  \right],
\end{equation}
having expressed $m_{\varphi_2}^2$ in terms of $m_{\varphi_1}^2$ according to eq.~\eqref{eq:m_phi2_vs_m_phi1}. Here, $A_i$ are coefficients expressed in terms of angles. For example,
\begin{equation} \label{Eq:trilin-A_H1}
\begin{aligned}
A_{H_1} =& \frac{3}{8} \cos^3 \theta_2 \bigg\lbrace 2 \cos \theta_1 \left[  \cos(2 \theta_2) + 5\right] \\  & \hspace{50pt}- \frac{2 \cos^2 \theta_2 \sin(2\beta - 3 \theta_1 ) - \cos(2\beta) \sin \theta_1 \left[ \cos(2 \theta_2) - 7 \right] }{\sin \beta \cos \beta}\bigg\rbrace.
\end{aligned}
\end{equation}
The trilinear coupling is shown in figure~\ref{Fig:C-III-a-trilinear}. Within the C-III-a model, either sign is possible, there is no simple correlation between the sign of the coupling and the parameters of the model. The SM-like limit, in terms of the gauge and Yukawa couplings, requires ${\cal R}^0_{11}=1$ and ${\cal R}^0_{1k}={\cal R}^0_{k1}=0$ ($k=2,3$). Indeed, expression (\ref{Eq:trilin-A_H1}) reduces to $A_{H_1} =3$ for $\theta_1=\theta_2=0$ and any $\beta$.

%%%%%%%%%%%%%%%%%%%%%%%%%%%%%%%%%%%%%%%%%%%%%%%%
\begin{figure}[htb]
\begin{center}
%\vspace*{-4mm}
\includegraphics[scale=0.35]{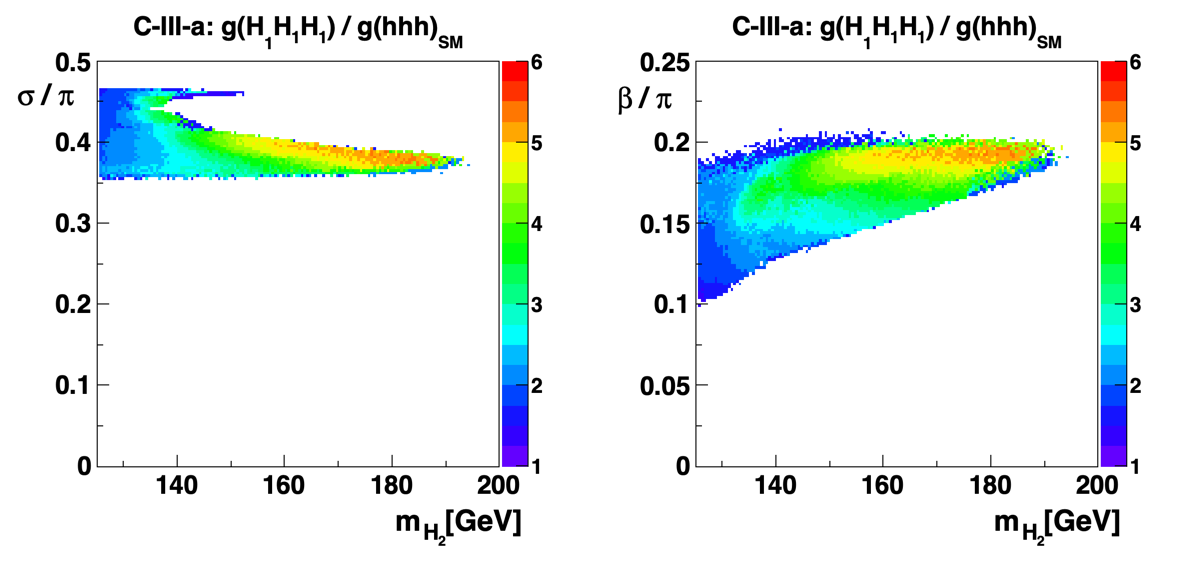}
\end{center}
\vspace*{-8mm}
\caption{Maximum values of the normalised trilinear coupling of the SM-like state $H_1$ as a function of $m_{H_2}$ and $\sigma$ (left) or $\beta$ (right), maximised over parameters not shown.}
\label{Fig:C-III-a-trilinear}
\end{figure}
%%%%%%%%%%%%%%%%%%%%%%%%%%%%%%%%%%%%%%%%%%%%%%%%

The form of the quartic self-interactions is similar to the trilinear one, but with different $A_i$ coefficients.

%%%%%%%%%%%%%%%%%%%%%%%%%%%%%%%%%%%%%%%%%%%%%%%%
\subsubsection{Astrophysical observables}
%%%%%%%%%%%%%%%%%%%%%%%%%%%%%%%%%%%%%%%%%%%%%%%%

We consider a standard cosmological model with a freeze-out scenario. The cold dark matter relic density along with the decay widths discussed above and other astrophysical observables are evaluated using $\mathsf{micrOMEGAs~5.2.7}$~\cite{Belanger:2008sj,Belanger:2013oya,Barducci:2016pcb}. The 't~Hooft-Feynman gauge is adopted, and switches are set to default values $\mathsf{VZdecay=VWdecay=1}$, specifying that 3-body final states will be computed for annihilation processes only. The $\mathsf{fast=-1}$ switch specifies that very accurate calculation is used.

We adopt the cold dark matter relic density value of $0.1200 \pm 0.0012$ taken from the PDG~\cite{Zyla:2020zbs}. The relic density parameter will be evaluated using a 3-$\sigma$ tolerance and assuming an additional ten per cent computational uncertainty,
\begin{equation} \label{Eq:Omega-value}
\begin{aligned}
\Omega h^2 &= 0.1200 \pm \sqrt{ \left( 0.1200 \times 0.1\right)^2  + (0.0012\,n)^2}\,,
\end{aligned}
\end{equation}
corresponding to the $[0.1075;\,0.1325]$ region. Results are presented in figure~\ref{Fig:Omega}. The relic density is found to fall quickly at DM masses beyond 50~GeV. 

%%%%%%%%%%%%%%%%%%%%%%%%%%%%%%%%%%%%%%%%%%%%%%%%
\begin{figure}[htb]
\begin{center}
%\vspace*{-4mm}
\includegraphics[scale=0.35]{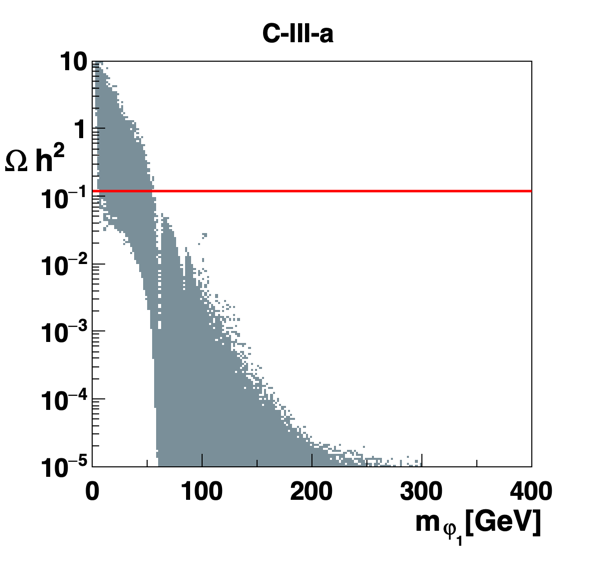}
\end{center}
\vspace*{-8mm}
\caption{Dark matter relic density for the C-III-a model. The region compatible with the observed DM relic density (red line) does not allow for masses above around $m_{H_1}/2$.}
\label{Fig:Omega}
\end{figure}
%%%%%%%%%%%%%%%%%%%%%%%%%%%%%%%%%%%%%%%%%%%%%%%%

The portal couplings $\varphi_1\varphi_1 H_i$ and $\varphi_1\varphi_1 H_i H_i$ play an important role for the Early Universe phenomenology. In the R-II-1a model we saw that the portal couplings increase very fast with high DM mass. Such high portal couplings imply a fast annihilation of DM, thus ruling out the possibility of obtaining the experimentally observed DM relic density for high dark matter masses. The absolute value of the trilinear portal couplings for C-III-a~\eqref{Eq.C_III_a_varphi1varphi1Hi} are illustrated in figure~\ref{Fig:C-III-a-portal}. The couplings can have either sign, but there is no simple correlation with the input parameters. 

%%%%%%%%%%%%%%%%%%%%%%%%%%%%%%%%%%%%%%%%%%%%%%%%
\vspace*{4mm}
\begin{figure}[htb]
\begin{center}
\includegraphics[scale=0.35]{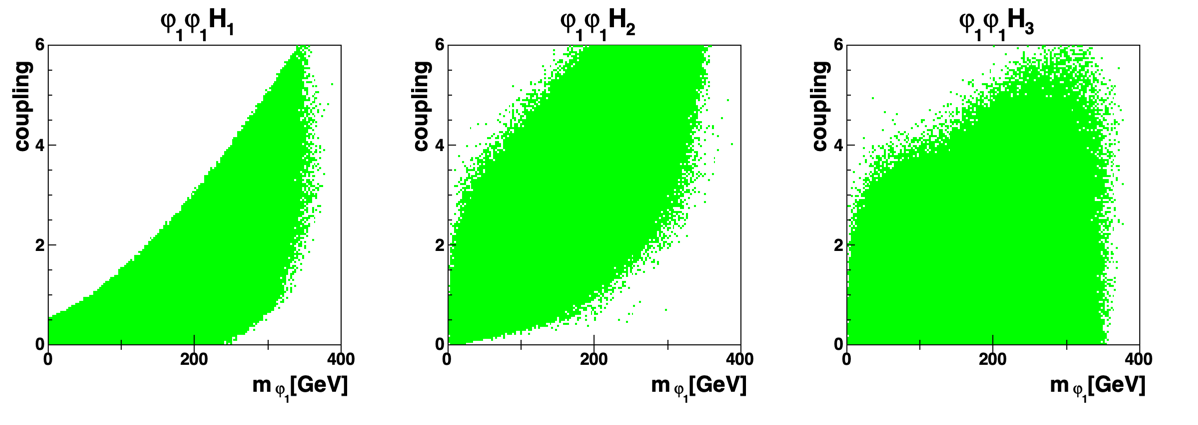}
\end{center}
\vspace*{-4mm}
\caption{Absolute values of the C-III-a portal couplings $g(\varphi_1 \varphi_1 H_i)$ as functions of the mass of the DM candidate $m_{\varphi_1}$.}
\label{Fig:C-III-a-portal}
\end{figure}
\vspace*{4mm}
%%%%%%%%%%%%%%%%%%%%%%%%%%%%%%%%%%%%%%%%%%%%%%%%

For a $\varphi_1$ scalar with mass above 300~GeV (Cut~1 allows for $m_{\varphi_1}^\text{max} \approx 500~\text{GeV}$ while Cut~2 shrinks the region to $m_{\varphi_1}^\text{max} \approx 400~\text{GeV}$) we get $\Omega h^2 \lsim \mathcal{O}(10^{-6})$. In this mass range the primary annihilation mechanisms are through the ${\varphi_1 \varphi_1 \to H_i H_j}$ channels. In the IDM the correct relic density, for high DM masses, is achieved due to a small portal coupling and near mass-degeneracy of the inert scalar sector. In section~\ref{Sec:C-III-a_Inert_Sector} we noted that it is not possible to have mass-degeneracy, $m_{\varphi_1} \approx m_{\varphi_2}$. There is always a mass gap. For heavy states, $m_{\varphi_1} \geq 300~\text{GeV}$, after applying the Cut~1 constraint a mass gap develops of around $m_{\varphi_2} \approx m_{\varphi_1} + 70~\text{GeV}$. The relevant processes for models with small, or vanishing, portal couplings would be diagrams with quartic vertices of the $SSVV$ type~\eqref{Eq:CIIIa_SSVV}. In the high-mass region, the $\Omega h^2$ parameter receives a contribution which grows as the difference of the squared inert-sector masses. Only for sufficiently low mass splittings between the inert-sector scalars can the correct relic density be reached. 

After separately applying each of the Cut~3 constraints to the parameter points satisfying Cut~1 and Cut~2, we found that the most severe constraint is the one due to the relic density. Less than one per cent of the Cut~1 and Cut~2-compatible points is satisfied after imposing the $\Omega h^2$ values. This is understandable after inspecting figure~\ref{Fig:Omega}. In fact, $\Omega h^2$ is not high enough in the region beyond $m_{H_1}/2$. However, in other models, as seen in figure~\ref{Fig:mass-ranges}, the surviving DM region (this is not an effect of only the relic density constraint) starts at values of $m_\text{DM} \approx 60~\text{GeV}$. In this region one would expect to see the most significant contribution from channels  $\text{DM}\, \text{DM}  \to \{b\,\bar{b},\, W^+ W^-, \, ZZ\}$. In contrast, in the C-III-a model the $\Omega h^2$ parameter drops below the experimental value for masses beyond about $m_{H_1}/2$. 
The most significant contribution, and the only adjustable (not fixed by the gauge coupling), comes from the portal couplings $\varphi_1 \varphi_1 H_i$. It turns out that the portal coupling to $H_2$ plays an important role in reducing the relic density for DM masses above some 50~GeV.

A less severe constraint comes from the direct detection analysis. An interesting aspect of the model is that the direct detection criteria are satisfied throughout the region \mbox{$m_{\varphi_1} \in [6,\, 360]~\text{GeV}$} and also at $m_{\varphi_1} \approx 1~\text{GeV}$. Two effects are responsible:
\begin{itemize}
\item Interference between different portal $\varphi_1 \varphi_1 H_i$ couplings;
\item The $H_i f\bar f$ couplings entering with both CP-even and CP-odd components;
\end{itemize}
The significance of these effects depends on the input parameters. We present cross sections relevant for direct detection in figure~\ref{Fig:Xenon1T}, comparing to the ``neutrino floor''. In practically the whole mass range there are parameter points at lower cross sections. A future improvement on the direct detection constraint is not obviously going to reduce the range of masses allowed by the model. Moreover, the cross section can be as low as $\sigma_\mathrm{SI} \approx 10^{-22}~\text{pb}$, which is way below the neutrino floor.

%%%%%%%%%%%%%%%%%%%%%%%%%%%%%%%%%%%%%%%%%%%%%%%%%%%%
\begin{figure}[h]
\begin{center}
%\vspace*{-4mm}
\includegraphics[scale=0.35]{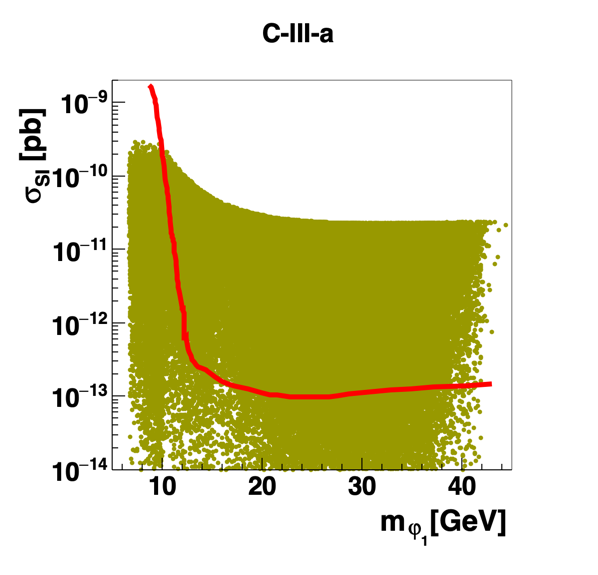}
\end{center}
\vspace*{-8mm}
\caption{ The spin-independent DM-nucleon cross section compatible with XENON1T~\cite{Aprile:2018dbl} data at 90\% C.L. The points represent Cut~3 satisfied cases. The red line corresponds to an approximate neutrino floor.}
\label{Fig:Xenon1T}
\end{figure}
%%%%%%%%%%%%%%%%%%%%%%%%%%%%%%%%%%%%%%%%%%%%%%%%%%%%

%%%%%%%%%%%%%%%%%%%%%%%%%%%%%%%%%%%%%%%%%%%%%%%%
\subsection{Cut 3 discussion}\label{sect:Cut_3}
%%%%%%%%%%%%%%%%%%%%%%%%%%%%%%%%%%%%%%%%%%%%%%%%

The LHC-related checks of Cut~3 are the least severe, satisfied by more than half of the parameter points surviving Cut~1 and Cut~2. When the DM candidate is sufficiently light, \mbox{$m_{\varphi_1} \leq m_{H_1}/2$}, decays of the SM-like Higgs particle into the dark sector, specifically the $H_1 \to \varphi_1 \varphi_1$ channel, play the most significant role. The high branching ratio of $H_1 \to \varphi_1 \varphi_1$ significantly impacts the total width of the SM-like Higgs particle, which is also constrained by Cut~3. One might expect that in the sub-$(m_{H_1}/2)$ region the decay of the SM-like Higgs particle into the invisible channel should be the most constraining one due to the need to tune the coupling. However, this is not the case, in this region both the relic density and direct detection constraints are even more demanding.

The model is described in terms of eight input parameters: three masses and five angles. For the purpose of discussion it is instructive to consider input in terms of just six masses, as was done for Cut~1 and Cut~2. First we apply each Cut~3 constraint separately, either the relic density constraint, or direct detection limits, or LHC related checks, over parameter points satisfying Cut~1 and Cut~2. 

There are no significant restrictions introduced on the charged masses. However, there are some restrictions introduced on the neutral inert sector masses. There is an upper limit $m_{\varphi_1} < 55~\text{GeV}$ and a lower limit $m_{\varphi_1} > 6~\text{GeV}$, both coming from the relic density constraint. The relic density checks allows also for $m_{\varphi_1} \approx 1~\text{GeV}$. The LHC checks restrict states lighter than $m_{\varphi_2} \approx 110~\text{GeV}$. These checks are very sensitive to the total width of the Higgs boson. Solutions with $m_{\varphi_2} < 110~\text{GeV}$ require $\Gamma_h > 0.2~\text{GeV}.$ There is a mass gap $m_{\varphi_2} \approx m_{\varphi_1} + 110~\text{GeV}$, for $m_{\varphi_1} < m_{H_1}/2$. The allowed masses of the neutral active sector are pushed away from the degenerate limit by both the relic density and the LHC constraints, so that $m_{H_3} > m_{H_2} + 20~\text{GeV}$.

%%%%%%%%%%%%%%%%%%%%%%%%%%%%%%%%%%%%%%%%%%%%%%%%
\begin{figure}[b!]
\begin{center}
\includegraphics[scale=0.3]{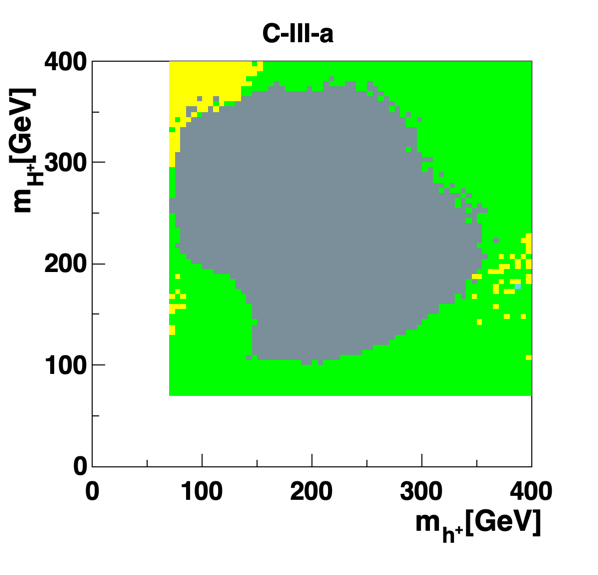}
\includegraphics[scale=0.3]{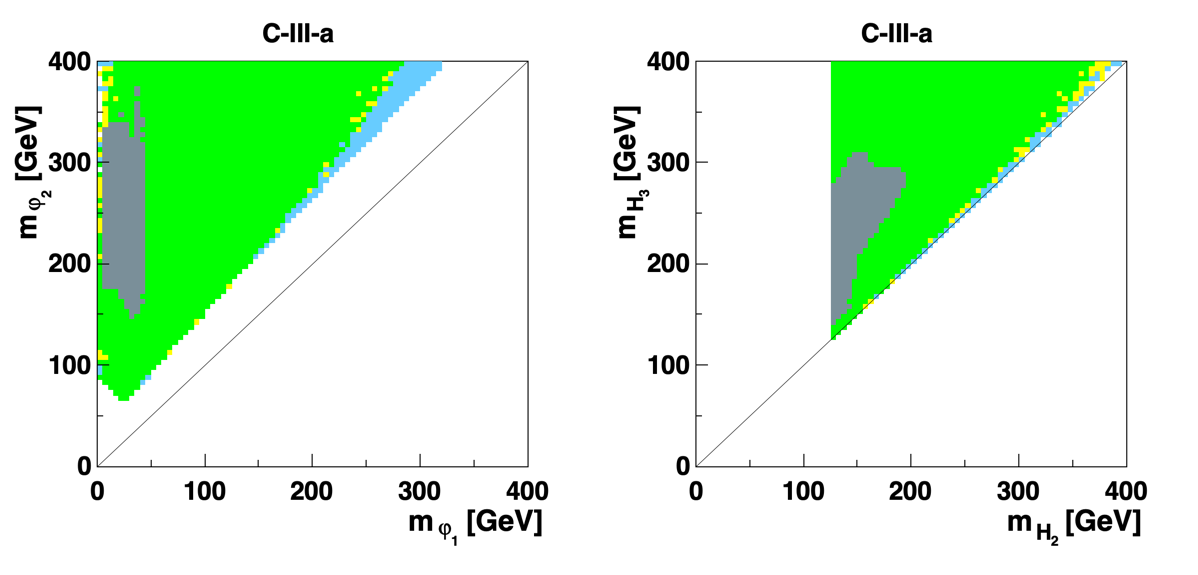}
\end{center}
\vspace*{-8mm}
\caption{Scatter plots of masses that satisfy different Cuts.
Top: the charged sector, $h^\pm$ and $H^\pm$.
Bottom left: the inert neutral sector, $\varphi_1$ and $\varphi_2$.
Bottom right: the active heavy neutral sector, $H_2$ and $H_3$.
The light-blue region satisfies Cut~1 and accommodates the $16\pi$ unitarity constraint.
The yellow region accommodates a 3-$\sigma$ tolerance with respect to Cut~2,  whereas in the green regions, the model is within the 2-$\sigma$ bound of these values.
The grey region is compatible with Cut~3.}
\label{Fig:C-III-a-masses-exp_constraint}
\end{figure}
%%%%%%%%%%%%%%%%%%%%%%%%%%%%%%%%%%%%%%%%%%%%%%%%

%%%%%%%%%%%%%%%%%%%%%%%%%%%%%%%%%%%%%%%%%%%%%%
{\renewcommand{\arraystretch}{1.25}
\begin{table}[b!] \small
\caption{Benchmark points and dominant decay modes. The ``$q$'' notation refers to a sum over the light quarks, $d$, $u$, $s$ and $c$, ``$\ell$'' refers to all charged leptons, and ``$\nu$'' to all neutrinos.}
\label{Table:benchmarks}
\begin{center}
\resizebox{14cm}{!}{
\begin{tabular}{|c|c|c|c|c|c|c|c|c|c|}
\hline\hline
Parameter & BP 1 & BP 2 & BP 3 & BP 4 & BP 5 & BP6 & BP7 & BP8 & BP9 \\
\hline
\hline
DM ($\varphi_1$) mass [GeV] & 6.85 & 11.55 & 16.24 & 20.82 & 25.50 & 30.36 & 35.13 & 39.73 & 44.24 \\ \hline
$\varphi_2$ mass [GeV] & 192.43 & 247.91 & 294.06 & 224.63 & 223.13 & 171.54 & 153.74 & 268.90 & 265.78 \\ \hline
$h^+$ mass [GeV] & 183.55 & 273.87 & 314.66 & 150.90 & 238.64 & 196.77 & 143.47 & 200.65 & 193.85 \\ \hline
$H^+$ mass [GeV] & 290.50 & 152.52 & 202.09 & 317.17 & 145.92 & 124.49 & 180.35 & 259.35 & 285.91 \\ \hline
$H_2$ mass [GeV] & 126.49 & 142.01 & 156.26 & 164.17 & 143.09 & 128.72 & 128.29 & 138.87 & 149.83 \\ \hline
$H_3$ mass [GeV] & 244.54 & 216.75 & 244.67 & 259.36 & 205.77 & 178.37 & 182.78 & 195.88 & 222.07 \\ \hline
$\sigma/\pi$ & 0.365 & 0.633 & -0.370 & -0.622 & -0.615 & -0.590 & 0.564 & -0.538 & -0.541 \\ \hline
$\beta/\pi$ & 0.167 & 0.146 & 0.160 & 0.191 & 0.139 & 0.128 & 0.138 & 0.152 & 0.150 \\ \hline
$\sigma_\text{SI}\, [10^{-11}~\text{pb}]$ & 9.23 & 1.55 & 1.45 & 0.01 & 0.10 & 1.65 & 1.23 & 0.67 & 3.09 \\ \hline\hline

$\varphi_2\to\varphi_1 H_1 $ [\%]  & 0.88 & 0.15 & 1.28 & 3.26 & 0.80 & 0.07 &  & 3.77 & 2.71 \\ \hline
$\varphi_2\to\varphi_1 H_2 $ [\%]  & 7.49 & 0.44 & 2.88 &  &  & 0.07 &  & 64.02 & 60.25 \\ \hline
$\varphi_2\to\varphi_1 H_3 $ [\%]  &  & 24.80 & 21.13 &  &  &  &  &  &  \\ \hline
$\varphi_2\to\varphi_1 Z $ [\%]  & 91.63 & 74.61 & 74.70 & 96.73 & 99.20 & 99.85 & 100 & 32.21 & 37.04 \\ \hline\hline

$h^+\to\varphi_1 H^+$ [\%] &  & 63.84 & 44.92 &  & 60.98 & 65.40 &  &  &  \\ \hline
$h^+\to\varphi_1 W^+$ [\%] & 100 & 36.16 & 55.08 & 100 & 39.02 & 34.60 & 100 & 100 & 100 \\ \hline\hline

$H^+\to h^+ \varphi_1$ [\%]  & 33.91 &  &  & 45.61 &  &  & 72.07 & 16.74 & 33.83 \\ \hline
$H^+\to H_1 W^+$ [\%]   & 2.26 &  &  & 3.10 &  &  &  & 2.25 & 2.50 \\ \hline
$H^+\to H_2 W^+$ [\%] & 15.19 &  &  & 9.34 &  &  &  & 10.73 & 10.55 \\ \hline
$H^+\to t\bar b$ [\%]   & 48.56 &  & 99.78 & 41.88 &  &  & 27.68 & 70.15 & 53.03 \\ \hline
$H^+\to q \bar q$ [\%]  & 0.08 & 29.32 & 0.17 & 0.06 & 29.49 & 30.14 &  & 0.10 & 0.08 \\ \hline
$H^+\to \nu \bar \ell$ [\%]  & 0.08 & 70.68 & 0.05 &  & 70.51 & 69.86 & 0.15 &  &  \\ \hline\hline

$H_2\to \varphi_1 \varphi_1$ [\%]  & 99.96 & 99.99 & 99.99 & 99.36 & 99.99 & 99.99 & 99.96 & 99.94 & 99.95 \\ \hline
$H_2\to W^+ W^-$ [\%]  &  &  &  & 0.60 &  &  &  &  &  \\ \hline
$H_2\to q \bar q$ [\%]  & 0.03 &  &  & 0.03 &  & 0.01 & 0.04 & 0.06 & 0.04 \\ \hline\hline

$H_3\to \varphi_1 \varphi_1$ [\%]  & 81.99 & 96.04 & 79.32 & 83.49 & 98.17 & 99.93 & 99.90 & 98.08 & 96.95 \\ \hline
$H_3 \to \varphi_1 \varphi_2$ [\%]  & 9.10 &  &  & 7.57 &  &  &  &  &  \\ \hline
$H_3 \to H_1 H_1$ [\%] &  &  &  & 0.08 &  &  &  &  &  \\ \hline
$H_3 \to H_1 Z$ [\%] & 1.20 &  & 15.82 & 2.57 &  &  &  &  & 0.01 \\ \hline
$H_3 \to H_2 Z$ [\%] & 7.67 &  &  & 0.40 &  &  &  &  &  \\ \hline
$H_3 \to W^+ W^-$ [\%]  &  & 2.64 & 3.18 & 4.10 & 1.26 & 0.04 & 0.08 & 1.44 & 2.17 \\ \hline
$H_3 \to Z Z$ [\%]  &  & 1.05 & 1.34 & 1.76 & 0.47 &  &  & 0.48 & 0.87 \\ \hline
$H_3 \to b \bar b$ [\%]  & 0.03 & 0.27 & 0.34 &  & 0.08 & 0.02 & 0.01 &  &  \\ \hline\hline
\end{tabular}
}
\end{center}
\end{table}}
%%%%%%%%%%%%%%%%%%%%%%%%%%%%%%%%%%%%%%%%%%%%%

Let us discuss cases when the Cut~3 constraints are introduced in pairs. When we assume $m_{\varphi_1} \lesssim 60~\text{GeV}$, as required by the relic density constraint, we find that there is a small difference between choosing different pairs of the Cut~3 constraints. A significant fraction of the Cut~1 and Cut~2-compatible parameter points is excluded in the charged sector when $\Omega h^2$ together with the LHC constraints are satisfied. This means that the full region of parameter space allowed by each of these two constraints separately only overlaps in a small region. The allowed region in the charged sector is practically reduced to what is shown in figure~\ref{Fig:C-III-a-masses-exp_constraint} (for all Cut~3 constraints). In the inert neutral sector a limit $m_{\varphi_2} > 100~\text{GeV}$ arises for any pair of constraints. Apart from that, any pair of constraints involving $\Omega h^2$ results in a bound $m_{\varphi_1} < 50~\text{GeV}$. Concerning the heavy active neutral sector, when both $\Omega h^2$ and LHC constraints are satisfied, an upper bound is introduced, $m_{H_3} < 300~\text{GeV}$.

In table~\ref{Table:benchmarks} we present some benchmarks. The more massive members of the inert doublet, $h^\pm$ and $\varphi_2$, are seen to predominantly decay to dark matter, $\varphi_1$, and a real gauge boson, $W^\pm$ or $Z$. Due to constraints coming from  Cut~3, see figure~\ref{Fig:C-III-a-masses-exp_constraint}, we note that there are lower bounds introduced on the masses of both $h^\pm$ and $\varphi_2$. Therefore, there are no co-annihilations into gauge bosons, nor can off-shell gauge bosons be produced.
The heavier non-inert neutral states, $H_2$ and $H_3$, decay almost exclusively to dark matter. This phenomenon is more pronounced for the $H_2$ scalar, for which \mbox{$\mathrm{Br}(H_2\to \varphi_1 \varphi_1)> 0.99$}. Also, the non-inert charged state has a significant branching ratio into members of the inert doublet, $H^+\to h^+\varphi_1$, in addition to those familiar from the 2HDM: $H^+\to \{ t\bar b,\,\nu\bar\ell \}$.

To sum up, the dominant decay channel for all of the scalars, except $H_1$, is into states with at least one dark matter candidate. Such processes would be accompanied by large missing transverse momentum in the detector. Depending on the parameters, this is only partially true for the active charged scalar, $H^\pm$. It would be interesting to further restrict the available parameter space of the charged state, specifically the $m_{H^\pm}\,-\,\beta$ plane based on decays into fermions~\cite{Arbey:2017gmh,ATLAS:2018gfm,CMS:2019bfg,CMS:2020imj,ATLAS:2021upq}. The acceptable parameter space of the C-III-a model could be reduced after applying additional constraints.

%%%%%%%%%%%%%%%%%%%%%%%%%%%%%%%%%%%%%%%%%%%%%%%%
\section{Concluding remarks}
\label{sect:conclude}
%%%%%%%%%%%%%%%%%%%%%%%%%%%%%%%%%%%%%%%%%%%%%%%%
We have extended our study of dark matter in 3HDMs based on $S_3$ symmetry from the model studied in Ref.~\cite{Khater:2021wcx}. There, we studied a model denoted R-II-1a with a zero vev for $h_1$ and the two other vevs real. In the present paper we study a model denoted C-III-a with the same vacuum structure as in R-II-1a, i.e., the vev of $h_1$ is still zero, but where now another vev is assumed to be  complex. In both cases we assume the coefficients of the potential to be real. The R-II-1a and C-III-a correspond to different regions of the parameter space of the $S_3$-symmetric potential  \cite{Emmanuel-Costa:2016vej}. The C-III-a model has the attractive feature of allowing for spontaneous CP violation and at the same time providing a dark-matter candidate.

The dark matter candidate, here referred to as $\varphi_1$, must have a mass below 50~GeV, which is lighter than the corresponding state in the familiar IDM. The reasons for this are mainly due to the possibility of suppressing the DM-DM-active neutral scalar couplings in C-III-a. We found that the acceptable DM mass range is $m_{\varphi_1} \in [6.5;\,44.5]~\text{GeV}$.

Compared to the familiar IDM, this model is very constrained. First of all, it is not possible to get correct relic density in the high-mass regime due to two effects: non-negligible portal couplings, which is the dominant effect, and a high mass splitting among the inert neutral states, of around 70~GeV. Moreover, heavy states with mass \mbox{$m_{\varphi_1} \gtrsim 500~\text{GeV}$} for the DM candidate are excluded after applying theoretical constraints (Cut~1). In the conventional lower-mass IDM region the relic density value is not satisfied due to portal couplings in the C-III-a model. The  sub-50 GeV region is accessible due to relatively low portal and scalar-fermion couplings. In the accompanying paper on the R-II-1a model~\cite{Khater:2021wcx}, the parameter space with a DM candidate with masses below 50 GeV was ruled out due to the lack of solutions satisfying simultaneously the relic density and direct detection constraints.

In the C-III-a model, the dark matter particle resides in an SU(2) doublet together with a heavier neutral scalar, $\varphi_2$, and a charged pair, $h^\pm$. These are unstable, and decay predominantly via the emission of an on-shell gauge boson, $\varphi_2\to \varphi_1 Z$ or $h^\pm\to\varphi_1 W^\pm$. 
The non-inert states have features similar to those of a Type-I CP-violating 2HDM. However, due to the constraints coming from the underlying $S_3$ symmetry, the scalar states are typically lighter than the corresponding 2HDM states. The charged states $H^\pm$ decay into a pair of fermions, either $tb$ or $\nu\ell$, or else to $h^\pm\varphi_1$. The neutral states, $H_2$ and $H_3$, predominantly decay to DM.

If the C-III-a model is realised in nature, it would be rather hard to detect it with current experiments. For the majority of the scalars the dominant decay channel is into states with at least one dark matter candidate. These decays would be accompanied by large missing transverse momentum in the detector. Moreover, there seems to be little hope of observing a signal based on DM direct detection. The spin-independent DM-nucleon cross section could be several orders of magnitudes lower than what would be probed by future dark matter direct detection experiments. In this work we applied a selected set of constraints on the C-III-a model, which are far from being exhaustive. It is beyond the scope of this paper to try to do a more comprehensive analysis. Our motivation is to show that the C-III-a model can in principle provide an interesting dark matter candidate. A more comprehensive study would definitely be justified in the future if there were experimental data pointing towards physics beyond the standard model of the kind we are outlining here.

\section*{Acknowledgements}

It is a pleasure to thank Igor Ivanov, Mikolaj Misiak and Alexander Pukhov for very useful discussions.
PO~is supported in part by the Research Council of Norway.
The work of AK and MNR was partially supported by Funda\c c\~ ao 
para  a  Ci\^ encia e a Tecnologia  (FCT, Portugal)  through  the  projects  
CFTP-FCT Unit UIDB/00777/2020 and UIDP/00777/2020, CERN/FIS-PAR/0008/2019 and 
PTDC/FIS-PAR/29436/2017  which are  partially  funded  through
POCTI  (FEDER),  COMPETE,  QREN  and  EU. Furthermore, the work of AK has been supported by the FCT PhD fellowship with reference UI/BD/150735/2020.
We also thank the University of Bergen  and
CFTP/IST/University of Lisbon, where collaboration visits took place. 

\appendix

%%%%%%%%%%%%%%%%%%%%%%%%%%%%%%%%%%%%%%%%%%%%%
\section{Determining C-III-a potential coefficients}
\label{App:C-III-a-lambdas}
\setcounter{equation}{0}
%%%%%%%%%%%%%%%%%%%%%%%%%%%%%%%%%%%%%%%%%%%%%
The model has eight $\lambda$'s of which $\lambda_4$ is fixed due to the minimisation condition~\eqref{eq:c-iii-a-lam4-lam7}, in terms of $\lambda_7$, $\sigma$ and $\beta$. We are thus left with seven free $\lambda$'s, which can be written in terms of seven mass-squared parameters. We note that masses of the $h^\pm$, $H^\pm$, $\varphi_1$, $\varphi_2$ states are expressed in terms of only four couplings: $\lambda_2$, $\lambda_3$, $\lambda_6$ and $\lambda_7$. We first discuss this sector. The remaining couplings can be written in terms of the masses involving also the $H_i$ states.

%%%%%%%%%%%%%%%%%%%%%%%%%%%%%%%%%%%%%%%%%%%%%
\subsection{The couplings $\{\lambda_2,\,\lambda_3,\,\lambda_6,\,\lambda_7\}$ vs masses of $\{h^\pm,\,H^\pm,\,\varphi_1,\,\varphi_2\}$}
\label{sect:lambdas_2_3_6_7}
%%%%%%%%%%%%%%%%%%%%%%%%%%%%%%%%%%%%%%%%%%%%%
With these four masses as input, together with $\hat w_2$, $\hat w_S$ and $\sigma$, one can determine the rotation angle $\gamma$ of eq.~(\ref{Eq:tan_2gamma}). This procedure leads to a quadratic equation, thus two sets of couplings for one and the same set of masses, and $\sigma$. In order to have more control on the input we shall rather sacrifice one mass, $m_{\varphi_2}$, replacing it by the rotation angle $\gamma$. This permits input of the basic masses, while leading to linear equations for the $\lambda$'s, thus unambiguous couplings. Eqs.~(\ref{Eq:C-III-c-masses-ch}) and (\ref{Eq:inert-masses}) can be solved in terms of $\lambda$'s, yielding:
\begin{subequations} \label{Eq:C-III-a-lambdas-a}
\begin{align}
\begin{split}
\lambda_2 &= \frac{m_{h^+}^2}{2 \hat{w}_2^2} - \frac{m_{H^+}^2 \hat{w}_S^2}{2v^2 \hat{w}_2^2}\\ &~~~ - \frac{ \left( m_{\varphi_1}^2 + m_{\varphi_2}^2\right)\left[ 2 + \cos (2\sigma) \right] + \left( m_{\varphi_2}^2 - m_{\varphi_1}^2 \right)\left[ 2 \cos (2\gamma) + \cos (2\gamma - 2\sigma) \right]}{12 \cos^2 \sigma \hat{w}_2^2},
\end{split}\\
\begin{split}
\lambda_3 &= -\frac{m_{h^+}^2}{2 \hat{w}_2^2} +  \frac{m_{H^+}^2 \hat{w}_S^2}{2v^2 \hat{w}_2^2} + \frac{\left( m_{\varphi_1}^2 + m_{\varphi_2}^2 \right)\sin \sigma - \left( m_{\varphi_2}^2 - m_{\varphi_1}^2 \right)\sin(2\gamma-\sigma)}{6 \hat{w}_2^2 \sin \sigma},
\end{split}\\
\lambda_6 & = -\frac{2m_{H^+}^2 }{v^2} + \frac{\left( m_{\varphi_1}^2 + m_{\varphi_2}^2 \right)\sin \sigma - \left( m_{\varphi_2}^2 - m_{\varphi_1}^2 \right)\sin(2\gamma-\sigma)}{12 \hat{w}_S^2 \cos^2 \sigma \sin \sigma},\\
\lambda_7 & = \frac{\left( m_{\varphi_1}^2 + m_{\varphi_2}^2 \right)\sin \sigma - \left( m_{\varphi_2}^2 - m_{\varphi_1}^2 \right)\sin(2\gamma-\sigma)}{24 \hat{w}_S^2 \cos^2\sigma \sin \sigma}. \label{Eq:C-III-a-lambda7}
\end{align}
\end{subequations}

Note that any expression
$\alpha(\gamma,\sigma) m^2_{\varphi_1}+\beta(\gamma,\sigma) m^2_{\varphi_2}$
can be expressed as
$A(\gamma,\sigma) m^2_{\varphi_1}+B(\gamma,\sigma) m^2_{\varphi_2}$ as long as
$\alpha+\beta g = A+B g$, with $g$ the ratio of the two coefficients in eq.~(\ref{eq:m_phi2_vs_m_phi1}).
Thus, we can write contributions to $\lambda$'s that involve $m^2_{\varphi_1}$ and $m^2_{\varphi_2}$ in many ways.

%%%%%%%%%%%%%%%%%%%%%%%%%%%%%%%%%%%%%%%%%%%%%
\subsection{The couplings $\{\lambda_1,\,\lambda_5,\,\lambda_8\}$ vs masses of $\{H_1,\,H_2,\,H_3\}$}
\label{sect:lambdas_1_5_8}
%%%%%%%%%%%%%%%%%%%%%%%%%%%%%%%%%%%%%%%%%%%%%

Eqs.~(\ref{Eq:C-III-a_NeutralActiveHBRot}, \ref{Eq:C-III-a_NeutralActiveHBRotR0Diag}) connects the remaining $\lambda$'s with the masses of the neutral non-inert sector. We find
\begin{subequations}\label{Eq:C-III-a_CouplingsInverted}
\begin{align}
\begin{split}
\lambda_1 & = \frac{m_{h^+}^2}{2 \hat{w}_2^2} - \frac{m_{H^+}^2 \hat{w}_S^2}{2v^2 \hat{w}_2^2} - \frac{\left( m_{\varphi_1}^2 + m_{\varphi_2}^2 \right)\sin \sigma - \left( m_{\varphi_2}^2 - m_{\varphi_1}^2 \right)\sin(2\gamma - \sigma)}{24 \hat{w}_2^2 \sin \sigma}\\
& ~~~ + \frac{1}{2 v^2 \hat{w}_2^2}\sum_{i=1}^3  \left( \mathcal{R}^0_{i1} \hat{w}_2 - \mathcal{R}^0_{i2} \hat{w}_S \right)^2 m_{H_i}^2,
\end{split}\raisetag{70pt}\\
\begin{split}
\lambda_5 & = \frac{2m_{H^+}^2 }{v^2} + \frac{\left( m_{\varphi_1}^2 + m_{\varphi_2}^2 \right)\sin \sigma - \left( m_{\varphi_2}^2 - m_{\varphi_1}^2 \right)\sin(2\gamma - \sigma)}{12 \hat{w}_S^2 \sin \sigma}\\
& ~~~  + \frac{1}{v^2 \hat{w}_2 \hat{w}_S} \sum_{i=1}^3  \left( \mathcal{R}^0_{i2} \hat{w}_2 + \mathcal{R}^0_{i1} \hat{w}_S \right)\left( \mathcal{R}^0_{i1} \hat{w}_2 - \mathcal{R}^0_{i2} \hat{w}_S \right) m_{H_i}^2,
\end{split}\\
\begin{split}
\lambda_8 & = -\frac{\left[\left( m_{\varphi_1}^2 + m_{\varphi_2}^2 \right)\sin \sigma - \left( m_{\varphi_2}^2 - m_{\varphi_1}^2 \right)\sin(2\gamma - \sigma)\right] \hat{w}_2^2}{24 \hat{w}_S^4 \sin \sigma}\\
& ~~~ +\frac{1}{2 v^2\hat w_S^2}\sum_i  (\mathcal{R}^0_{i2}\hat w_2+\mathcal{R}^0_{i1}\hat w_S)^2 m_{H_i}^2 .
\end{split}
\end{align}
\end{subequations}
In addition, the diagonalisation matrix $\mathcal{R}^0$ should satisfy
\begin{subequations}\label{Eq:C-III-a_ActiveSectorAnglesConstraints}
\begin{align}
\sum_i \mathcal{R}^0_{i1}\mathcal{R}^0_{i3} m_{H_i}^2&=0, \label{Eq:non-inert-constraint1}\\
\sum_i \mathcal{R}^0_{i2}\mathcal{R}^0_{i3} m_{H_i}^2&=v^2\sin(2\sigma)\lambda_7, \label{Eq:non-inert-constraint2}\\\
\sum_i \left(\mathcal{R}^0_{i3}\right)^2m_{H_i}^2&=2v^2\sin^2\sigma \lambda_7. \label{Eq:non-inert-constraint3}
\end{align}
\end{subequations}
We note that the two last constraints, eqs.~(\ref{Eq:non-inert-constraint2}) and (\ref{Eq:non-inert-constraint3}), relate the mass scale of the non-inert neutral sector $m_{H_i}^2$ with that of the inert-sector neutral states $m_{\varphi_i}$ via $\lambda_7$ given by eq.~(\ref{Eq:C-III-a-lambda7}). This way the squares of masses $m_{H_i}^2$ can be expressed as:
\begin{subequations}
\begin{align}
m_{H_1}^2 & = \Phi \left( \sin \theta_1 \cot \theta_2 + \tan \sigma\right), \label{Eq:CIIIa_mH12_theta}\\
m_{H_2}^2 & = \Phi \left( - \sin \theta_1 \tan \theta_2 + \frac{\cos \theta_1 \cot \theta_3}{\cos \theta_2}   + \tan \sigma\right),\\
m_{H_3}^2 & = \Phi \left( - \sin \theta_1 \tan \theta_2 - \frac{\cos \theta_1 \tan \theta_3}{\cos \theta_2} + \tan \sigma\right),
\end{align}
\end{subequations}
with 
\begin{equation}
\Phi = \frac{m_{\varphi_1}^2 v^2 \sin \left( 2 \gamma - 2 \sigma \right) \sin \sigma }{ 3 \hat{w}_S^2 \left[ \sin \left( 2 \gamma - 3 \sigma \right) + 2 \sin\left( 2 \gamma - \sigma \right) - \sin \sigma \right]},
\end{equation}
where $\Phi$ can be both negative and positive. 

With the mass-squared parameters some of the conditions~\eqref{Eq:C-III-a_ActiveSectorAnglesConstraints} can be re-written:
\begin{subequations}
\begin{align}
\eqref{Eq:non-inert-constraint2}:&\quad \Phi = v^2\sin(2\sigma)\lambda_7,\\
\eqref{Eq:non-inert-constraint3}:&\quad \Phi \tan \sigma = 2v^2\sin^2\sigma \lambda_7.
\end{align}
\end{subequations}

In a scan over parameters, it is obviously desirable to keep $m_{H_1}$ fixed at the experimental value. This can be achieved within this framework. The constraints~\eqref{Eq:C-III-a_ActiveSectorAnglesConstraints} allow for taking one mass and two angles, or two masses and one angle, or three masses as input, in addition to those discussed in appendix~\ref{sect:lambdas_2_3_6_7}. From eq.~\eqref{Eq:CIIIa_mH12_theta}, $m_{H_1}^2 = f\left(\theta_1, \theta_2\right)$, it follows that it is not possible to use an arbitrary combination of masses and angles as input. In our scan we use $\theta_2$ and $\theta_3$ as input along with the $m_{H_1}$ state corresponding to the SM-like Higgs boson.

%%%%%%%%%%%%%%%%%%%%%%%%%%%%%%%%%%%%%%%%%%%%%
\section{Scalar-scalar couplings of C-III-a}
\label{App:Scalar_Couplings_CIIIa}
\setcounter{equation}{0}
%%%%%%%%%%%%%%%%%%%%%%%%%%%%%%%%%%%%%%%%%%%%%

The scalar-scalar couplings are presented with the symmetry factor, but without the overall coefficient ``$-i$". We denote the ``correct" couplings as $g_{\dots} = -i g\left({\dots}\right)$. We shall abbreviate $\mathrm{c}_\theta\equiv\cos\theta$, and $\mathrm{s}_\theta\equiv\sin\theta$, and $\mathrm{t}_\theta\equiv\tan\theta$ for any argument $\theta$.

For simplicity, we introduce a permutation function, which for trilinear couplings takes the form
\begin{equation}
P_{\overline{m} n o}(i,j,k) = \sum\limits_{\substack{t_\alpha \in \{i,j,k\}}} A^\ast_{mt_1} A_{nt_2} A_{ot_3}, 
\end{equation}
where the $A$'s are coefficients of the field expansions, defined in eq.~\eqref{Eq:C-III-a_Expanded_Mass_Eigenstates}.
Furthermore, the indices $\{i,j,k\}$ are carried by the fields $H_i H_j H_k$, and the barred index $\overline{m}$ indicates which of the $A$'s are conjugated.
As an example, the permutation function $P_{\overline{2} 2 S}(i,j,k) $ which enters the $H_i H_j H_k$ vertex is
\begin{equation}
P_{\overline{2} 2 S}(i,j,k) = A^\ast_{2i} (A_{2j} A_{Sk}+A_{2k}A_{Sj})
+A^\ast_{2j}(A_{2i}A_{Sk}+A_{2k}A_{Si})
+A^\ast_{2k}(A_{2i}A_{Sj}+A_{2j}A_{Si})
\end{equation}
Based on the number of the involved $H_i$ scalars in a vertex, the permutation function $P$ can also be of length two, $P_{\overline{m} n}(i,j)$, and four, $P_{\overline{m} \overline{n} o p}(i,j,k,\ell)$. For example,
\begin{equation}
P_{2S}(i,j)=A_{2i}A_{Sj}+A_{Si}A_{2j}.
\end{equation}
Note that the order of $m$, $n$ and $o$ is arbitrary, 
\begin{equation}
P_{\overline{m} n o}(i,j,k) = P_{n \overline{m} o}(i,j,k) = P_{n o \overline{m}}(i,j,k), \text{ and interchange of }n\leftrightarrow o,
\end{equation}
and that
\begin{equation}
\left(P_{\overline{m} n o}(i,j,k)\right)^\ast = P_{m \overline{n} \overline{o}}(i,j,k).
\end{equation}

Furthermore, in the interest of obtaining more compact expressions, we here suppress the fact that $\lambda_4$ is proportional to $\lambda_7$~\eqref{eq:c-iii-a-lam4-lam7}.

The trilinear couplings involving the neutral fields are:
\begin{subequations}
\begin{align}
\begin{split}
g\left( H_i H_j H_k \right) &= v \Big\{\frac{1}{2}\left( \lambda_1 + \lambda_3 \right) \mathrm{s}_\beta 
P_{\overline{2}22}(i,j,k)\\
&\hspace{30pt} - \frac{1}{4}\lambda_4 e^{i \sigma} \{ \mathrm{c}_\beta P_{\overline{2} 22}(i,j,k) 
+ \mathrm{s}_\beta \left[ P_{\overline{S} 22}(i,j,k) + 2 P_{\overline{S2}2}(i,j,k) \right] \} \\
&\hspace{30pt} + \frac{1}{4} \left( \lambda_5 + \lambda_6 \right) \left( \mathrm{c}_\beta P_{\overline{2} 2S}(i,j,k)  
+ \mathrm{s}_\beta P_{2S\overline{S}}(i,j,k)  \right)\\
&\hspace{30pt} + \frac{1}{2}\lambda_7 e^{2 i \sigma} \left[ \mathrm{c}_\beta P_{\overline{S} 22}(i,j,k)  
+ \mathrm{s}_\beta P_{2\overline{SS}}(i,j,k)\right]\\
&\hspace{30pt} + \frac{1}{2}\lambda_8 \mathrm{c}_\beta P_{S\overline{SS}}(i,j,k) + \mathrm{h.c.} \Big\}\label{Eq.C_III_a_HiHjHk},
\end{split}\\
\begin{split}
g\left( \varphi_1 \varphi_1 H_i \right) &= v \Big\{ \lambda_1 \mathrm{s}_\beta A_{2i} - \lambda_2  \left( 1-e^{-2i \left( \gamma - \sigma \right)} \right)\mathrm{s}_\beta A_{2i} 
+ \lambda_3  e^{2i\left(\sigma-\gamma\right)}\mathrm{s}_\beta A_{2i}\\
&\hspace{30pt}+\frac{1}{2}\lambda_4 \left[ e^{i\sigma}\left( 2+e^{-2i\gamma}\right)\mathrm{c}_\beta  A_{2i} + e^{-i \sigma}\left( 2 + e^{-2i\left(\gamma-\sigma\right)}\right)\mathrm{s}_\beta A_{Si}  \right]\\
&\hspace{30pt}+ \frac{1}{2}\left( \lambda_5 + \lambda_6 \right) \mathrm{c}_\beta A_{Si} + \lambda_7 e^{-2i \gamma}\mathrm{c}_\beta A_{Si} + \mathrm{h.c.}  \Big\},
\end{split}\label{Eq.C_III_a_varphi1varphi1Hi}\\
\begin{split}
g\left( \varphi_1 \varphi_2 H_i \right) 
&= v \Big\{ -i e^{-2i\gamma}[\left( \lambda_2 + \lambda_3 \right)e^{2i\sigma}\mathrm{s}_\beta A_{2i} 
+ \frac{1}{2}  \lambda_4 e^{i\sigma}\left( \mathrm{c}_\beta A_{2i} 
+ \mathrm{s}_\beta A_{Si} \right)\\
&\hspace{75pt} + \lambda_7 \mathrm{c}_\beta A_{Si}] + \mathrm{h.c.}\Big\},
\end{split}\label{Eq.C_III_a_varphi1varphi2Hi}\\
\begin{split}
g\left( \varphi_2 \varphi_2 H_i \right) &= v \Big\{ \lambda_1 \mathrm{s}_\beta A_{2i} - \lambda_2  \left( 1+e^{-2i \left( \gamma - \sigma \right)} \right)\mathrm{s}_\beta A_{2i} - \lambda_3  e^{2i\left(\sigma-\gamma\right)}\mathrm{s}_\beta A_{2i}\\
&\hspace{30pt}+\frac{1}{2}\lambda_4 \left[ e^{i\sigma}\left( 2-e^{-2i\gamma}\right)\mathrm{c}_\beta  A_{2i} + e^{-i \sigma}\left( 2 - e^{-2i\left(\gamma-\sigma\right)}\right)\mathrm{s}_\beta A_{Si}  \right]\\
&\hspace{30pt}+ \frac{1}{2}\left( \lambda_5 + \lambda_6 \right) \mathrm{c}_\beta A_{Si} - \lambda_7 e^{-2i \gamma}\mathrm{c}_\beta A_{Si} + \mathrm{h.c.}  \Big\}\label{Eq.C_III_a_varphi2varphi2Hi}.
\end{split}
\end{align} 
\end{subequations}

The trilinear couplings involving the charged fields are:
\begin{subequations}
\begin{align}
\begin{split}
g\left( \varphi_1 h^+ H^- \right) &= v \Big\{ \frac{1}{2}\lambda_2 \left( 1 - e^{2i\left(\gamma - \sigma \right)} \right)\mathrm{s}_{2\beta} - \frac{1}{2}\lambda_3 \left( 1 + e^{2i\left(\gamma - \sigma \right)} \right)\mathrm{s}_{2\beta}\\
&\hspace{30pt} + \lambda_4 \left[ -\frac{1}{2}e^{-i\sigma}\left( 1 + e^{2i\gamma}\right)\mathrm{c}_\beta^2 + e^{i\gamma} \mathrm{c}_{\gamma-\sigma} \mathrm{s}_\beta^2 \right]\\
&\hspace{30pt} + \frac{1}{4}\lambda_6 \mathrm{s}_{2\beta} + \frac{1}{2}\lambda_7 e^{2i\gamma}\mathrm{s}_{2\beta}  \Big\},
\end{split}\\
\begin{split}
g\left( \varphi_2 h^+ H^- \right) &= v \Big\{ -i\frac{1}{2}\lambda_2 \left( 1 + e^{2i\left(\gamma - \sigma \right)} \right)\mathrm{s}_{2\beta} + i\frac{1}{2}\lambda_3 \left( 1 - e^{2i\left(\gamma - \sigma \right)} \right)\mathrm{s}_{2\beta}\\
&\hspace{30pt} + \lambda_4 \left( e^{i\left( \gamma - \sigma\right)}\mathrm{s}_\gamma \mathrm{c}_\beta^2 - e^{i\gamma} \mathrm{s}_{\gamma - \sigma} \mathrm{s}_\beta^2 \right)\\
&\hspace{30pt} -i\frac{1}{4}\lambda_6 \mathrm{s}_{2\beta} + i\frac{1}{2}\lambda_7 e^{2i\gamma}\mathrm{s}_{2\beta}  \Big\},
\end{split}\\
g \left( H_i h^\pm h^\mp \right) &= v \Big\{ \left( \lambda_1 - \lambda_3 \right)\mathrm{s}_\beta A_{2i} + \frac{1}{2} \lambda_4 e^{i \sigma} \left( \mathrm{c}_\beta A_{2i} + \mathrm{s}_\beta A_{Si}^\ast \right) + \frac{1}{2} \lambda_5 \mathrm{c}_\beta A_{Si} + \mathrm{h.c.} \Big\},\\
\begin{split}
g\left( H_i H^\pm H^\mp  \right) &= v \Big\{ \left( \lambda_1 + \lambda_3 \right)\mathrm{c}_\beta^2 \mathrm{s}_\beta A_{2i} - \frac{1}{2}\lambda_4 e^{i \sigma} \mathrm{c}_\beta \left[ \mathrm{c}_{2\beta}A_{2i} + \frac{1}{2}\mathrm{s}_{2\beta}A_{Si}^\ast - \mathrm{s}_\beta^2 A_{2i}^\ast \right]\\
&\hspace{30pt} +\frac{1}{2}\lambda_5 \left( \mathrm{c}_\beta^3 A_{Si} + \mathrm{s}_\beta^3 A_{2i} \right) - \frac{1}{4}\lambda_6 \mathrm{s}_{2\beta} \left( \mathrm{c}_\beta A_{2i} + \mathrm{s}_\beta A_{Si} \right) \\
&\hspace{30pt} - \frac{1}{2}\lambda_7 e^{2i\sigma}\mathrm{s}_{2\beta}\left( \mathrm{c}_\beta A_{2i} + \mathrm{s}_\beta A_{Si}^\ast \right) + \lambda_8 \mathrm{c}_\beta \mathrm{s}_\beta^2 A_{Si} + \mathrm{h.c.} \Big\}.
\end{split}
\end{align} 
\end{subequations}
Note that couplings involving charged fields of different sectors, $g(\varphi_1 h^+ H^-)$ and $g(\varphi_2 h^+ H^-)$, are complex. For opposite charges, the couplings are obtained by complex conjugation.

The quartic couplings involving only the neutral fields are:
\begin{subequations}
\begin{align}
g \left( \varphi_1 \varphi_1 \varphi_1 \varphi_1 \right) &= g \left( \varphi_2 \varphi_2 \varphi_2 \varphi_2 \right) = 6 \left( \lambda_1 + \lambda_3 \right)\label{Eq.C_III_a_4varphi},\\
g \left( \varphi_1 \varphi_1 \varphi_2 \varphi_2 \right) &= 2 \left( \lambda_1 + \lambda_3 \right),\\
\begin{split}
g \left( \varphi_1 \varphi_1 H_i H_j \right) 
&= \frac{1}{2}\lambda_1 P_{2\overline{2}}(i,j) + \frac{1}{2} \lambda_2 \left[ e^{2i\left(\sigma - \gamma \right)}P_{22}(i,j)  
- P_{2\overline{2}}(i,j)\right]
\\
&\hspace{15pt}+ \frac{1}{2}\lambda_3 e^{2i\left( \sigma - \gamma \right)} P_{22}(i,j)  + \lambda_4 \left( \frac{1}{2}e^{i\left( \sigma - 2 \gamma \right)}P_{2S}(i,j)  
+ e^{i \sigma} P_{2\overline{S}}(i,j)  \right)\\
&\hspace{15pt} + \frac{1}{4} \left( \lambda_5 + \lambda_6 \right) P_{S\overline{S}}(i,j) + \frac{1}{2} \lambda_7 e^{2i \gamma }P_{\overline{SS}}(i,j)  + \mathrm{h.c.},
\end{split}\\
\begin{split}
g \left( \varphi_1 \varphi_2 H_i H_j \right) 
&= -\frac{i}{2}\left(\lambda_2 + \lambda_3 \right)e^{2i \left(\sigma - \gamma\right)}P_{22}(i,j) 
- \frac{i}{2}\lambda_4 e^{i \left( \sigma - 2\gamma \right)}P_{2S}(i,j)\\
&\hspace{15pt} + \frac{i}{2}\lambda_7 e^{2i\gamma}P_{\overline{SS}}(i,j) +  \mathrm{h.c.},
\end{split}\\
\begin{split}
g \left( \varphi_2 \varphi_2 H_i H_j \right) 
&= \frac{1}{2}\lambda_1 P_{2\overline{2}}(i,j) - \frac{1}{2} \lambda_2 \left[ e^{2i\left(\sigma - \gamma \right)}P_{22}(i,j)  
+ P_{2\overline{2}}(i,j)\right]\\
&\hspace{15pt}- \frac{1}{2}\lambda_3 e^{2i\left( \sigma - \gamma \right)} P_{22}(i,j)  - \lambda_4 \left( \frac{1}{2}e^{i\left( \sigma - 2 \gamma \right)}P_{2S}(i,j)  
- e^{i \sigma} P_{2\overline{S}}(i,j)  \right)\\
&\hspace{15pt} + \frac{1}{4} \left( \lambda_5 + \lambda_6 \right) P_{S\overline{S}}(i,j) - \frac{1}{2} \lambda_7 e^{2i \gamma }P_{\overline{SS}}(i,j) + \mathrm{h.c.},
\end{split}\\
\begin{split}
g \left( H_i H_j H_k H_l \right) 
&= \frac{1}{8}\left( \lambda_1 + \lambda_3 \right) P_{22\overline{22}}(i,j,k,l) 
- \frac{1}{4} \lambda_4 e^{i \sigma} P_{22\overline{2S}}(i,j,k,l) \\
&\hspace{15pt} + \frac{1}{8}\left( \lambda_5 + \lambda_6 \right) P_{2\overline{2}S\overline{S}}(i,j,k,l)  
+ \frac{1}{4} \lambda_7 e^{2i\sigma} P_{22\overline{SS}}(i,j,k,l) \\
&\hspace{15pt} + \frac{1}{8} \lambda_8 P_{SS\overline{SS}}(i,j,k,l)  + \mathrm{h.c.}\label{Eq.C_III_a_HiHjHkHl}
\end{split}
\end{align} 
\end{subequations}
The last quartic coupling has been expressed compactly in terms of {\it four} indices, at least two of which have to be identical. For example, the $H_1H_1H_2H_2$ coupling is obtained with $i=j=1$ and $k=l=2$, without any further combinatorial factors.

The quartic couplings involving both neutral and charged fields are:
\begin{subequations}
\begin{align}
g \left( \varphi_1 \varphi_1 h^\pm h^\mp \right) &= g \left( \varphi_2 \varphi_2 h^\pm h^\mp \right) = 2 \left( \lambda_1 + \lambda_3 \right),\\
g \left( \varphi_1 \varphi_1 H^\pm H^\mp \right) &= g \left( \varphi_2 \varphi_2 H^\pm H^\mp \right) = 2 \left( \lambda_1 - \lambda_3 \right)\mathrm{c}_\beta^2 - \lambda_4 \mathrm{c}_\sigma \mathrm{s}_{2\beta} + \lambda_5 \mathrm{s}_\beta^2,\\
\begin{split}
g \left( \varphi_1 H_i h^+ H^- \right) 
&= \lambda_2 \mathrm{c}_\beta \left( A_{2i} - e^{2i\left( \gamma - \sigma \right)}A_{2i}^\ast \right) 
-\lambda_3 \mathrm{c}_\beta \left( A_{2i} + e^{2i\left( \gamma - \sigma \right)}A_{2i}^\ast \right)\\
&\hspace{15pt} - \frac{1}{2}\lambda_4 \left[ \mathrm{c}_\beta \left( e^{-i \sigma}A_{Si} + e^{i\left( 2\gamma - \sigma \right)}A_{Si}^\ast \right) - \mathrm{s}_\beta \left( e^{i \sigma}A_{2i} + e^{i\left( 2\gamma - \sigma \right)}A_{2i}^\ast \right) \right]\\
&\hspace{15pt} + \frac{1}{2} \lambda_6 \mathrm{s}_\beta A_{Si} 
+ \lambda_7 e^{2i \gamma}\mathrm{s}_\beta A_{Si}^\ast,
\end{split}\\
\begin{split}
g \left( \varphi_2 H_i h^+ H^- \right) 
&= -i\lambda_2 \mathrm{c}_\beta \left( A_{2i} + e^{2i\left( \gamma - \sigma \right)}A_{2i}^\ast \right) 
+ i\lambda_3 \mathrm{c}_\beta \left( A_{2i} - e^{2i\left( \gamma - \sigma \right)}A_{2i}^\ast \right)\\
&\hspace{15pt} + \frac{i}{2}\lambda_4 \left[ \mathrm{c}_\beta \left( e^{-i \sigma}A_{Si} - e^{i\left(2\gamma - \sigma \right)}A_{Si}^\ast \right) - \mathrm{s}_\beta \left( e^{i \sigma}A_{2i} - e^{i\left(2\gamma - \sigma \right)}A_{2i}^\ast \right) \right]\\
&\hspace{15pt} - \frac{i}{2} \lambda_6 \mathrm{s}_\beta A_{Si} 
+ i\lambda_7 e^{2i \gamma}\mathrm{s}_\beta A_{Si}^\ast,
\end{split}\\
g \left( H_i H_j h^\pm h^\mp \right) 
&= \frac{1}{2}\left( \lambda_1 - \lambda_3 \right)P_{2\overline{2}}(i,j) 
+ \frac{1}{2} \lambda_4 e^{i \sigma} P_{2\overline{S}}(i,j)
+ \frac{1}{4} \lambda_5 P_{S\overline{S}}(i,j) + \mathrm{h.c.},\\
\begin{split}
g \left( H_i H_j H^\pm H^\mp \right) 
&= \frac{1}{2}\left( \lambda_1 + \lambda_3 \right) \mathrm{c}_\beta^2 P_{2\overline{2}}(i,j) 
- \frac{1}{2}\lambda_4 e^{i\sigma}\mathrm{c}_\beta 
\left[ \mathrm{c}_\beta P_{2\overline{S}}(i,j) - \mathrm{s}_\beta P_{2\overline{2}}(i,j) \right]\\
&\hspace{15pt} + \frac{1}{4} \lambda_5 \left[ \mathrm{c}_\beta^2 P_{S\overline{S}}(i,j) 
+ \mathrm{s}_\beta^2 P_{2\overline{2}}(i,j) \right] 
- \frac{1}{4}\lambda_6 \mathrm{s}_{2\beta}P_{\overline{2}S}(i,j) \\
&\hspace{15pt} - \frac{1}{2}\lambda_7 e^{2i\sigma} \mathrm{s}_{2\beta} P_{2\overline{S}}(i,j) 
+\frac{1}{2}\lambda_8 \mathrm{s}_\beta^2 P_{S\overline{S}}(i,j)  + \mathrm{h.c.}
\end{split}
\end{align} 
\end{subequations}
The quartic couplings involving only the charged fields are:
\begin{subequations}
\begin{align}
g\left( h^\pm h^\pm h^\mp h^\mp \right) & = 4 \left( \lambda_1 + \lambda_3 \right),\label{Eq:hphphmhm}\\
g\left( H^\pm H^\pm H^\mp H^\mp \right) & = 4 \left[ \left( \lambda_1 + \lambda_3\right)\mathrm{c}_\beta^4  + 2 \lambda_4 \mathrm{c}_\sigma \mathrm{c}_\beta^3 \mathrm{s}_\beta + \frac{1}{4} \left( \lambda_a -4 \lambda_7 \mathrm{s}_\sigma^2 \right)\mathrm{s}_{2\beta}^2 + \lambda_8 \mathrm{s}_\beta^4 \right],\\
g\left( h^+ h^+ H^- H^- \right) & = 4 e^{2i(\gamma-\sigma)}\left[ \left( \lambda_2 + \lambda_3 \right)\mathrm{c}_\beta^2 - \frac{1}{2}e^{i \sigma} \lambda_4 \mathrm{s}_{2\beta} + e^{2 i \sigma} \lambda_7 \mathrm{s}_\beta^2 \right],\\
g\left( h^\pm h^\mp H^\pm H^\mp \right) & = \left[ 2 \left( \lambda_1 - \lambda_2 \right)\mathrm{c}_\beta^2 - 2\lambda_4 \mathrm{c}_\sigma \mathrm{s}_{2\beta} + \left( \lambda_5 + \lambda_6 \right) \mathrm{s}_\beta^2 \right].
\end{align} 
\end{subequations}

%%%%%%%%%%%%%%%%%%%%%%%%%%%%%%%%%%%%%%%%%%%%%
\section{Supplementary equations}
\label{App:Additional}
\setcounter{equation}{0}
%%%%%%%%%%%%%%%%%%%%%%%%%%%%%%%%%%%%%%%%%%%%%

%%%%%%%%%%%%%%%%%%%%%%%%%%%%%%%%%%%%%%%%%%%%%
\subsection{\boldmath$V$ and \boldmath$U$ matrices }\label{App:Peskin_Takeuchi_Rot}
%%%%%%%%%%%%%%%%%%%%%%%%%%%%%%%%%%%%%%%%%%%%%
From Refs.~\cite{Grimus:2007if,Grimus:2008nb} we determine the $V$ and $U$ matrices\footnote{Note that ``$U$'' here should not be confused with the electroweak precision observable ``$U$''.}  for C-III-a:
\begin{subequations}
\begin{equation}
 \begin{pmatrix}
e^{i \sigma} \left( i G^0 + \sum_{i=1}^3 \mathcal{R}^0_{i1} H_i \right) \\
\sum_{i=1}^3 \left( \mathcal{R}_{i2}^0 + i \mathcal{R}_{i3}^0 \right) H_i \\
e^{i \gamma} \left(\varphi_1 + i \varphi_2\right) \\
\end{pmatrix} = V \begin{pmatrix}
G^0 \\
H_1 \\
H_2 \\
H_3 \\
\varphi_1 \\
\varphi_2 \\
\end{pmatrix},
\end{equation}
with
\begin{equation}
V = \begin{pmatrix}
i e^{i\sigma} & \mathcal{R}_{11}^0 e^{i\sigma} & \mathcal{R}_{21}^0 e^{i\sigma}  & \mathcal{R}_{31}^0 e^{i\sigma} & 0 & 0 \\
0 & \left( \mathcal{R}_{12}^0 + i \mathcal{R}_{13}^0 \right) & \left( \mathcal{R}_{22}^0 + i \mathcal{R}_{23}^0\right) & \left( \mathcal{R}_{32}^0 + i \mathcal{R}_{33}^0 \right) & 0 & 0 \\
0 & 0 & 0 & 0 &  e^{i\gamma} & i e^{i\gamma}
\end{pmatrix}, 
\end{equation}
and
\begin{equation}
 \begin{pmatrix}
G^+ \\
H^+ \\
h^+ \\
\end{pmatrix} = U \begin{pmatrix}
G^+ \\
H^+ \\
h^+ \\
\end{pmatrix}, \text{ with } U=\mathrm{diag}\left(e^{i\sigma},\,1,\,e^{i\gamma}\right).
\end{equation}
\end{subequations}

%%%%%%%%%%%%%%%%%%%%%%%%%%%%%%%%%%%%%%%%%%%%%
\subsection{Higgs decays}\label{App:DiGamma}
%%%%%%%%%%%%%%%%%%%%%%%%%%%%%%%%%%%%%%%%%%%%%

We assume that the normalised Lagrangian for $H_1$ is given by:
\begin{equation}
\begin{aligned}
\mathcal{L}_\text{int}^\prime  =\,&-\frac{ m_f}{v}  C_{\bar{f}fH_1}^S \bar{f}fH_1 -i\frac{ m_d}{v}  C_{\bar{d}d H_1}^P \bar{d} \gamma_5 dH_1 + i\frac{ m_u}{v}  C_{\bar{u}u H_1}^P \bar{u} \gamma_5 u H_1\\
&+  g m_W C_{W^+W^-H_1} W_\mu^+ W^{\mu -} H_1  - \frac{2   m_{\varphi^\pm_i}^2}{v} C_{ \varphi^+_i \varphi^-_i h}  \varphi^+_i \varphi^-_i H_1,
\end{aligned}
\end{equation}
where $C$'s are the couplings normalised to those of the SM,

The rate for the two-gluon decay at the leading order is~\cite{Wilczek:1977zn,Georgi:1977gs,Ellis:1979jy,Rizzo:1979mf}
\begin{equation}
\Gamma\left(H_1 \rightarrow g g \right) =\frac{\alpha_S^2 m_{H_1}^3}{128 \pi^3 v^2}
\left( \left| \sum_f C_{\bar{f}fH_1}^S \mathcal{F}_{1/2}^S(\tau_f) \right|^2 + \left| \sum_f C_{\bar{f}fH_1}^P \mathcal{F}_{1/2}^P(\tau_f) \right|^2 \right),
\end{equation}
where $\alpha_S$ is the strong coupling constant. The decay width of this process can be enhanced or diminished with respect to the SM case. Such behaviour is caused by an additional factor for the amplitude and the fact that there is an additional contribution from the CP-odd part.

The diphoton decay one-loop width is known~\cite{Ellis:1975ap, Shifman:1979eb, Gunion:1989we}:
\begin{equation}\label{Eq:h_gammagamma}
\begin{aligned}
\Gamma\left(H_1 \rightarrow \gamma \gamma\right)&=\frac{\alpha^2 m_{H_1}^3}{256 \pi^3 v^2}\Bigg(\bigg|\sum_f  Q_f^2 N_c C_{\bar{f}fH_1}^S \mathcal{F}_{1 / 2}^S\left(\tau_{f}\right) + C_{W^+W^-H_1} \mathcal{F}_1\left(\tau_{W^\pm}\right) \\ 
&\hspace{35pt} + \sum_{\varphi^\pm_i}C_{ \varphi^+_i \varphi^-_i H_1} \mathcal{F}_0\left(\tau_{\varphi^\pm}\right) \bigg|^{2}  + \bigg|\sum_f Q_f^2 N_c C_{\bar{f}fH_1}^P \mathcal{F}_{1 / 2}^P\left(\tau_{f}\right)\bigg|^{2} \Bigg),
\end{aligned}
\end{equation}
where $\alpha$ is the fine-structure constant, $Q_f$ is the electric charge of the fermion, $N_c = 3\,(1)$ for quarks (leptons).

The one-loop spin-dependent functions are
\begin{subequations}
\begin{align}
\mathcal{F}_{1} &=2+3 \tau+3 \tau(2-\tau) f(\tau), \\
\mathcal{F}_{1 / 2}^i &= \left\{ \begin{aligned}
& -2 \tau[1+(1-\tau) f(\tau)], & i = S,\\
& -2 \tau f(\tau), & i = P,
\end{aligned}\right. \\
\mathcal{F}_{0} &=\tau[1-\tau f(\tau)],
\end{align}
\end{subequations}
where
\begin{equation}
\tau_i = \frac{4 m_i^2}{m_{H_1}^2},
\end{equation}
and 
\begin{equation}
f(\tau)=\left\{
\begin{aligned}
&\arcsin ^{2}\left(\frac{1}{\sqrt{\tau}}\right), & \tau \geq 1, \\
&-\frac{1}{4}\left[\ln \left(\frac{1+\sqrt{1-\tau}}{1-\sqrt{1-\tau}}\right)-i \pi\right]^{2}, & \tau<1.
\end{aligned}\right.
\end{equation}

The decay width of $H_1$ into a pair of scalars $\varphi_i$ is given by
\begin{equation}\label{Eq.DecayWidth_SSS}
\Gamma\left( H_1 \to \varphi_i \varphi_j \right) = \frac{2-\delta_{ij}}{32 \pi m_{H_1}^3} \left| g_{H_1 \varphi_i \varphi_j} \right|^2 \sqrt{\left[ m_{H_1}^2 - \left( m_{\varphi_i} + m_{\varphi_j} \right)^2 \right]\left[ m_{H_1}^2 - \left( m_{\varphi_i} - m_{\varphi_j} \right)^2 \right]},
\end{equation}
with a symmetry factor $(2-\delta_{ij})$, where $\delta_{ij}$ is the Kronecker delta. After applying the cuts it was found that $m_{\varphi_2} > m_{H_1}$, and hence the invisible decay rate simplifies to
\begin{equation}
\Gamma\left( H_1 \to \varphi_1 \varphi_1 \right) = \frac{1}{32 \pi m_{H_1}^2} \left| g_{H_1 \varphi_1 \varphi_1} \right|^2 \sqrt{ m_{H_1}^2 -  4m_{\varphi_1}^2}.
\end{equation}

\bibliographystyle{JHEP}

\bibliography{ref}

\end{document}